\documentclass[11pt]{article}
\DeclareMathAlphabet{\scr}{U}{rsfs}{m}{n}

\pdfoutput=1
\usepackage{latexsym}
\usepackage{epsfig}
\usepackage[mathscr]{eucal}
\usepackage{amsfonts}
\usepackage{amscd}
\usepackage{cite}
\usepackage{array}
\usepackage{bbold}
\usepackage{amssymb}
\usepackage{colordvi}
\usepackage[centertags]{amsmath}
\usepackage{enumerate}
\usepackage{graphicx}
\usepackage{booktabs}
\usepackage{theorem}
\usepackage[footnotesize]{caption}
\usepackage{soul}
\usepackage{url}
\usepackage{mcite}
\usepackage{slashed}
\usepackage[dvipsnames]{xcolor}
\usepackage{bbm}
\usepackage[utf8]{inputenc}
\usepackage{hyperref}
\hypersetup{colorlinks=true, linkcolor=teal, urlcolor=blue, citecolor=red}

\usepackage{booktabs} 

\allowdisplaybreaks{}

\newcommand{\newc}{\newcommand}

\newc{\ol}{\overline}
\newc{\wt}{\widetilde}
\newc{\bs}{\boldsymbol}
\newc{\m}{\mathcal}
\newc{\la}{\langle}
\newc{\ra}{\rangle}

\newcommand{\beq}{\begin{eqnarray}}
\newcommand{\eeq}{\end{eqnarray}}
\newcommand{\bpmatrix}{\begin{pmatrix}}
\newcommand{\epmatrix}{\end{pmatrix}}

\renewcommand{\ol}{\text{1l}}

\renewcommand{\eqref}[1]{Eq.~(\ref{#1})}

\newcommand{\bc}{\begin{center}}
\newcommand{\ec}{\end{center}}

\definecolor{britishracinggreen}{rgb}{0.0, 0.26, 0.15}

\usepackage[dvipsnames]{xcolor}

\newenvironment{Eqnarray}%
     {\arraycolsep 0.14em\begin{eqnarray}}{\end{eqnarray}}
\newcommand{\ba}{\begin{Eqnarray}}
\newcommand{\ea}{\end{Eqnarray}}
\newcommand{\be}{\begin{equation}}
\newcommand{\ee}{\end{equation}}

\begin{document}

\title{
\vspace*{-3.7cm}
\phantom{h} \hfill\mbox{\small }
\\[1cm]
\textbf{Impact of SM parameters and of the vacua of the Higgs potential in gravitational waves detection \\[4mm]}}

\date{}
\author{
Felipe F. Freitas$^{1, \, 2\,}$\footnote{E-mail:
\texttt{felipefreitas@ua.pt}} ,
Gabriel Louren\c co$^{3,\,}$\footnote{E-mail:
\texttt{gabaslourenco@gmail.com}} ,
Ant\'onio P.~Morais$^{1, \, 2\,}$\footnote{E-mail:
\texttt{aapmorais@ua.pt}} ,
Andr\'e Nunes$^{3\,}$\footnote{E-mail:
\texttt{andre.martins.nunes@gmail.com}} ,\\
Jo\~ao Ol\'{\i}via$^{3\,}$\footnote{E-mail:
\texttt{joao.s.olivia@gmail.com}} ,
Roman Pasechnik$^{4\,}$\footnote{E-mail:
\texttt{Roman.Pasechnik@thep.lu.se}} ,
Rui Santos$^{3, \, 5\,}$\footnote{E-mail:
\texttt{rasantos@fc.ul.pt}} ,
Jo\~ao Viana$^{3\,}$\footnote{E-mail:
\texttt{jfvvchico@hotmail.com}},
\\[5mm]
{\small\it
$^1$ Departamento de F\'isica, Universidade de Aveiro,} \\
{\small \it  Campus de Santiago, 3810-183 Aveiro, Portugal} \\[3mm]
{\small\it
$^2$ Centre for Research and Development in Mathematics and Applications (CIDMA),} \\
{\small \it   3810-183 Aveiro, Portugal} \\[3mm]
{\small\it
$^3$Centro de F\'{\i}sica Te\'{o}rica e Computacional,
Faculdade de Ci\^{e}ncias,} \\
{\small \it    Universidade de Lisboa, Campo Grande, Edif\'{\i}cio C8
1749-016 Lisboa, Portugal} \\[3mm]
{\small\it
$^4$Department of Astronomy and Theoretical Physics, Lund University,}\\
{\small\it Lund University, 221 00 Lund, Sweden}\\[3mm]
{\small\it
$^5$ISEL -
 Instituto Superior de Engenharia de Lisboa,} \\
{\small \it   Instituto Polit\'ecnico de Lisboa
1959-007 Lisboa, Portugal} \\[3mm]
}

\maketitle

\begin{abstract}
In this work we discuss two different phases of a complex singlet extension of the Standard Model (SM) together with an extension that also includes new fermion fields, in particular, a Majoron model equipped with an inverse seesaw mechanism. All considered scenarios contain a global $\mathrm{U}(1)$ symmetry and allow for first-order phase transitions while only two of them are strong enough to favour the detection of primordial gravitational waves (GWs) in planned experiments such as LISA. In particular, this is shown to be possible in the singlet extension with a non vanishing real VEV at zero temperature and also in the model with extra fermions. 
In the singlet extension with no additional fermions, the detection of GWs strongly depends on the $\mathrm{U}(1)$ symmetry breaking pattern of the scalar potential at zero temperature. We study for the first time the impact of the precision in the determination of the SM parameters on the strength of the GWs spectrum. It turns out that the variation of the SM parameters such as the Higgs boson mass and top quark Yukawa coupling in their allowed experimental ranges has a notable impact on GWs detectability prospects.
\end{abstract}

\thispagestyle{empty}
\vfill
\newpage
\setcounter{page}{1}

\maketitle

\section{Introduction}
\label{Sec:Intro}

As often happens in physics, there is an apparently strange connection between particle physics and gravitational waves (GWs). Assuming that the Higgs potential at zero temperature is a result of a strong first order phase transition (FOPT) in the Higgs vacuum that occurred in the early Universe, signs of that transition could appear today in the form of primordial GWs and, under particular circumstances, could be detected in a not too distant future~\cite{Witten:1984rs, Hogan:1986qda, PhysRevLett.65.3080, Kamionkowski:1993fg}. A strong EW FOPT is considered to be an important prerequisite for the generation of baryon asymmetry in the early Universe as one of the Sakharov conditions~\cite{Sakharov:1967dj}. It is well known that the Higgs potential of the Standard Model (SM) cannot provide a FOPT and an extension, even minimal, is needed in order to include this new feature in the model. One of the most simple extensions of the Higgs potential, with the addition of a singlet field (with Isospin and Hypercharge zero), is sufficient to trigger a strong enough FOPT needed for EW baryogenesis~\cite{Espinosa:2007qk}. The implications for the detection of GWs in the case of the singlet-extended Higgs potential were first discussed in~Ref.~\cite{Ashoorioon:2009nf} while an extension with an arbitrary number of singlets was studied in~Ref.~\cite{Kakizaki:2015wua} (other simple extension like two-Higgs doublet models were discussed in~\cite{Dorsch:2013wja, Basler:2016obg, Goncalves:2021egx}). These and other works study a connection between the values of the parameters of the scalar potential, the characteristics of the FOPTs (their duration and latent heat) and properties of the associated primordial GW spectrum such as its peak amplitude and frequency. Once the potential is fixed it is possible to search for the regions of the parameter space that would give rise to potentially detectable GW signatures.

The relation between the parameters of the Higgs potential and the strength of the GW spectrum is prone to large instabilities that can be traced back to the multidimensional nature of the field content as well as to numerical instabilities in the calculation of the bounce action and its derivative. In fact, the phase transition happens at a very specific point of the parameter space and field configuration and it is possible that a point extremely close to that one will not undergo a phase transition. This fact leads to the following question: would this instability be also reflected on the sensitivity of the results to the SM parameters and if so to what extent? And there is one more question to be asked: how do the codes presently used deal with such instabilities? Hence, one of the goals of this paper is to understand the effect of the precision in the measurement of the SM parameters in the characteristics of the GW spectra. 

Another interesting point that we will address is if different realisations of the spontaneous symmetry breaking of a given model may lead to a significant difference in the strength of the GW signal. In fact, we may ask ourselves if when we study a specific model, say the complex singlet extension of the SM, and allow for two different patterns of the symmetry breaking, how disparate the corresponding GW spectra can be. Moreover, if the differences are indeed significant, how does the addition of new particles affect the GW spectrum? All these issues are discussed by analysing the FOPTs and their GW signatures in a simple extension of the SM featuring an additional complex scalar singlet in two different phases, as well as in the scope of the Majoron model with inverse seesaw mechanism described in~Ref.~\cite{Addazi:2019dqt}.

The paper is organized as follows. In section~\ref{Sec:scalar} we briefly present the potential and the vacuum structure of the model and in section~\ref{Sec:GWs} we discuss the relation between GWs and FOPT. In section~\ref{sec:GWSc1} we discuss the detection of GW in the case where the singlet  acquires a vacuum expectation value (VEV) at zero temperature followed in section~\ref{sec:SMpar} by a discussion on the dependence of GW spectra on the SM parameters for the same model. We then move to a discussion of the impact of the dark sector on the strength of the GWs signatures in section~\ref{sec:dark}. Finally, in section~\ref{sec:3Sc} we compare three different scenarios and summarise our findings in section~\ref{sec:Conc}.

\section{The models}
\label{Sec:scalar}

In this section, we present the models that will serve as a basis for the discussion of the GW spectrum dependence on the SM parameters. We will focus on three different scenarios that have in common the scalar sector which consists of a simple extension of the SM potential with an extra complex gauge singlet $\sigma$ (with hypercharge $Y = 0$). This potential features a global (and, in general, softly-broken) $U(1)$ symmetry, such that the Higgs doublet $\Phi$ and complex EW singlet $\sigma$ can be given a non-trivial charge under this symmetry. Although having the same scalar potential, in the first two scenarios no other new particles are added while the last scenario is motivated by an inverse seesaw mechanism for neutrino mass generation where the new scalar singlet plays the role of a Majoron~\cite{Addazi:2019dqt,Gonzalez-Garcia:1988okv,Chikashige:1980ui,Schechter:1981cv}. In the latter case, the global $U(1)$ symmetry of the potential is extended to the lepton sector and is thus called the lepton-number $U(1)_L$ symmetry, and we adopt this notation in all three scenarios. Due to the presence of a $U(1)_L$-charged complex scalar field $\sigma$ the global $U(1)_L$ symmetry is considered to be both explicitly and spontaneously broken. Such a pattern of symmetry reduction is accomplished via a mass term of the singlet $\sigma$ and by means of the generation of a VEV in the singlet field $\sigma$. If no new fermions are added, the model is always CP-conserving and we take all parameters of the potential real also when new fermions are added, focusing therefore only on a CP-conserving model, for simplicity.

In any of the scenarios to be discussed below, the scalar potential is written as follows
\begin{eqnarray}
\mathcal{V}_0(\Phi,\sigma) & = & \mu_\Phi^2 \Phi^\dag \Phi + \lambda_\Phi (\Phi^\dag \Phi)^2 
+ \mu_\sigma^2 \sigma^\dag \sigma + \lambda_\sigma (\sigma^\dag \sigma)^2 \nonumber \\
& & + \lambda_{\Phi \sigma} \Phi^\dag \Phi \sigma^\dag \sigma + 
\Big(\frac12 \mu_b^2 \sigma^2 + {\rm h.c.}\Big) \, ,
\label{V0-tot}
\end{eqnarray}
with $\Phi$ and $\sigma$ given by
\begin{align}
\begin{aligned}
\Phi &= \frac{1}{\sqrt{2}} 
\begin{pmatrix} 
G + i G' \\ 
\phi_h + h + i \eta 
\end{pmatrix}\,,
\end{aligned} 
\quad
\begin{aligned}
 \sigma = \dfrac{1}{\sqrt{2}}(\phi_\sigma + \sigma_R + i \sigma_I ) \,,
\end{aligned} 
\end{align}
where $h$, $\eta$, $G$, $G'$, $\sigma_R$, $\sigma_I$ are the real scalars. The Higgs field, $h$, is a quantum fluctuation about the classical mean-field $\phi_h$, which in the zero temperature limit approaches the corresponding SM Higgs VEV $\phi_{h}(T=0)\equiv v_{h} = 246$ GeV. 

As previously stated, we study three distinct scenarios, where in the first two only a complex singlet field is added to the SM, while the third scenario corresponds to a low-scale inverse seesaw with Majoron -- a popular scenario for neutrino mass generation as detailed in~Ref.~\cite{Addazi:2019dqt} and references therein. A brief description of the considered scenarios is as follows:
\begin{itemize}
\item \textbf{Scenario 1} -- in this case only the doublet acquires a VEV whereas both the real and imaginary components of the gauge singlet have zero VEVs at $T=0$ \textit{i.e.}~$\phi_{\sigma}(T=0)=0$. At finite temperatures though, the real part may fluctuate around a non-zero $\phi_\sigma(T)$. This implies that the model may feature two possible dark matter (DM) candidates. One was always a DM particle since the beginning of the Universe while the other, for certain none zero temperatures, featured a temperature dependent mixing with the neutral component from the doublet, vanishing as $T\to 0$. The $U(1)_L \to \mathbb{Z}_2$ soft breaking term in~\eqref{V0-tot} provides a pseudo-Goldstone mass to the imaginary part of the EW singlet field.
\item \textbf{Scenario 2} -- in this case both the doublet and the real component of the gauge singlet acquire VEVs at $T=0$, that is, $\phi_{h,\sigma}(T=0)\equiv v_{h,\sigma}$. One of the CP-even scalar states is identified with the SM-like Higgs boson with a mass of $125$ GeV. The second scalar, that mixes with the SM-like Higgs boson, can be either heavier or lighter than the 125 GeV Higgs boson candidate in this case. The soft breaking term in the potential explicitly breaks $U(1)_L \to \mathbb{Z}_2$ providing a pseudo-Goldstone mass to the imaginary part of the field $\sigma_I$ known in many contexts as a Majoron.
\item \textbf{Scenario 3} -- From the point of view of this work, Scenario 3 can be seen as an extension of Scenario 2. The scalar potential is exactly the same but right-handed neutrinos are introduced in the context of an inverse seesaw mechanism. The details of the Majoron model as well as its constraints are discussed in~Refs.~\cite{Addazi:2019dqt,Gonzalez-Garcia:1988okv,Chikashige:1980ui,Schechter:1981cv}.
%
\end{itemize}

Let us now describe the first two scenarios in more detail. We first note that such scenarios are just two different phases of the same potential at zero temperature.
This means that there are conditions that are the same in both cases. The conditions for the potential to be bounded from below read 
\begin{equation}
\lambda_\Phi > 0, \quad \lambda_\sigma > 0, \quad \lambda_{\Phi \sigma} > - 2 \sqrt{\lambda_\Phi \lambda_\sigma} \,,
\end{equation}
and will be imposed in our numerical calculations. Also, in our analysis we impose a conservative perturbativity bound on the quartic couplings, $\lambda_{\Phi \sigma}, \lambda_\sigma < 2 \pi$.

In scenario 1, the mass spectrum at zero temperature is just the one of the SM with two new dark scalars. The SM-like Higgs boson has a mass, $m_h\simeq 125$ GeV and emerges entirely from the doublet. This in turn also means that the Higgs couplings to the remaining SM particles are not modified. The DM candidates only couple to the Higgs boson via the portal coupling $\lambda_{\Phi \sigma}$ which can be constrained by measurements of the invisible Higgs decay as well as by direct and indirect DM detection data. In the scalar sector the mass spectrum is given by
\begin{eqnarray}
&& m_{h}^2 = 2\lambda_\Phi v_h^2 \,, \quad m_{D1}^2 = \mu_\sigma^2 + \mu_b^2 + \frac{\lambda_{\Phi \sigma} v_h^2}{2}\,, 
\quad  m_{D2}^2 = \mu_\sigma^2 - \mu_b^2 + \frac{\lambda_{\Phi \sigma} v_h^2}{2} \,,
\label{mA}
\end{eqnarray}
for the SM Higgs boson and for the two DM candidates, $D1$ and $D2$, respectively. Which of these two is the stable one at zero temperature depends
on sign of $\mu_b^2$ parameter. Indeed,
\begin{equation}
m_{D2}^2 - m_{D1}^2= - 2 \mu_b^2 \,,
\end{equation}
which means that $D2$ is the DM particle if $\mu_b^2 > 0$, while $D1$ is the DM candidate if $\mu_b^2 < 0$. 
In practice, it may be possible to consider also a scenario with one stable and one metastable DM candidates.

In scenario 2, the main difference is in the particle spectrum. Now the CP-even component of the singlet mixes with the CP-even component from the doublet and only one DM candidate remains. 
The masses of the CP-even states can be written as
\begin{eqnarray}
m_{h_1,h_2}^2=\lambda_\Phi v_h^2 + \lambda_\sigma v_\sigma^2 \mp 
\frac{\lambda_\sigma v_\sigma^2 - \lambda_\Phi v_h^2}{\cos 2\theta} \,,
\label{eq:mh12}
\end{eqnarray}
in terms of $h$-$\sigma_R$ mixing angle $\theta$, while the the DM candidate gets a pseudo-Goldstone mass,
\begin{eqnarray}
m_D^2\equiv m_{\sigma_I}^2 = - 2\mu_b^2 \,, \qquad \mu_b^2 < 0 \,.
\label{eq:mD}
\end{eqnarray}

Finally, scenario 3 is exactly the same as scenario 2 in what concerns the scalar sector and the only difference resides in the fermion content of the model. In particular, three families of right-handed neutrinos $\nu_{1,2,3}^c$ carrying lepton number $L(\nu^c) = -1$ and three families of singlet fermions $S_{1,2,3}$ with the opposite lepton number, \textit{i.e.}~$L(S) = 1$, are introduced such that one can write two additional Yukawa interactions, one of them of the Dirac-like tying together the Higgs, lepton doublets and right-handed neutrinos while the other coupling is of the Majorana-like and ties the two singlet fermions with the complex singlet, $Y_{\sigma_i} S S \sigma$ (for this reason, $\sigma$ is dubbed Majoron in this model) which is invariant under the lepton number $\mathrm{U}_\mathrm{L}$ symmetry provided that $L(\sigma) = -2$. Note that as long as the singlet $\sigma$ develops a VEV, a Majorona mass term of the form $\mu_i S_i S_i$ is induced, with
\begin{equation}
	\mu_i = \dfrac{Y_{\sigma_i} }{\sqrt{2}} v_\sigma\,,
\end{equation}
where, for simplicity of illustration, we assume a flavour diagonal basis. The lepton number symmetry also allows a mass term of form $M_i \nu_i^c S_i$, which is also considered in our numerical analysis. Since the singlet VEV is expected to be generated not far from the EW scale, the model features a low-scale type-I seesaw mechanism. 

\section{Gravitational waves from FOPTs}
\label{Sec:GWs}

In this section, we will define the physical quantities relevant for understanding the characteristics of the GW signals originating from EW FOPTs in the early Universe. A detailed knowledge of the effective scalar potential at finite temperatures $V_{\rm eff}(\phi_{\alpha};T)$ is important in order to obtain the key parameters of the primordial GWs power spectrum. To the one-loop order, the effective potential takes the form~\cite{Quiros:1999jp,Curtin:2016urg},
\begin{equation}
V_{\rm eff}(T) = V_0 + V^{(1)}_{\rm CW} + \Delta V(T) + V_{\rm ct}\,,
\label{eff-pot}
\end{equation}
in terms of $V_0$ and $V^{(1)}_{\rm CW}$ being the tree-level (classical) part and one loop Coleman-Weinberg (CW) potential, respectively, and the counterterm potential $V_{\rm ct}$, while finite-temperature corrections are denoted as $\Delta V(T)$.

The one-loop zero-temperature effective potential is given by the standard formula~\cite{Coleman:1973jx} (in the $\overline{MS}$ scheme and in the Landau gauge)
\be
 V^{(1)}_{\rm CW} =\frac{1}{64 \pi^2}\sum_{a}n_a m_a^4(h,\phi)\left[\log\frac{m_a^2(h,\phi)}{\mu^2}-C_a\right], \label{eq:one-loop}
\ee
where $n_a$ counts the number of degrees of freedom and for a particle of spin $s_a$ is given by
\be
n_a=(-1)^{2s_a} Q_a N_a (2s_a+1),\notag
\ee
where $N_a$ stands for the number of colours and $Q_a=1,2$ for neutral/charged particles. $m_a(h,\phi)$ correspond to (tree-level) field-dependent masses.

One-loop thermal corrections are given by \cite{Quiros:1999jp}
\begin{equation}
\Delta V(T) = \frac{T^4}{2 \pi^2} \left\{ \sum_{b} n_b J_B\left[\frac{m_i^2(\phi_\alpha)}{T^2}\right] - \sum_{f} n_f J_F\left[\frac{m_i^2(\phi_\alpha)}{T^2}\right] \right\}\,,
\label{finite_T_correction}
\end{equation} 
where $J_B$ and $J_F$ are the thermal integrals for bosons and fermions, respectively, provided by
\begin{align} \label{eq:JBJF}
J_{B/F}(y^2) = \int_0^\infty d x \, x^2 \log\left( 1 \mp \exp [ - \sqrt{x^2 + y^2}] \right)\,.
\end{align}
For the first non-trivial order of the thermal expansion $\sim(m/T)^2$, $\Delta V(T)$ can be approximated as
\begin{align}
\Delta V^{(1)}(T)|_{\rm L.O.} = \frac{T^2}{24} \left\{ {\rm Tr}\left[ M_{\alpha\beta}^2(\phi_\alpha) \right] + 
\sum_{i=W,Z,\gamma} n_i m_i^2(\phi_\alpha) + 
\sum_{i = \mathrm{f}_i} \frac{n_i}{2} m_i^2(\phi_\alpha) \right\} \,,
\label{eq:DV_LO}
\end{align}
where in the last sum we consider all the fermions in the considered models consisting of three generations of quarks and charged leptons for scenarios 1 and 2, as well as six heavy neutrinos in what concerns scenario 3. The first term in \eqref{eq:DV_LO} denotes the trace of the field-dependent scalar Hessian matrix $M_{\alpha\beta}^2(\phi_\alpha)$, which is a basis invariant quantity. In the case of the discussed models we have used Eqs.~(\ref{eq:mh12}) and (\ref{eq:mD}) upon replacing the VEVs by their classical field configurations $v_h \to \phi_h$ and $v_\sigma \to \phi_\sigma$. This means that the leading thermal corrections only affect the quadratic terms (in mean-fields) of the scalar potential, preserving the shape of $V_{0}$ and affecting only the masses of the scalar fields. The $n_i$ coefficients in \eqref{eq:DV_LO} represent the number of d.o.f for a given particle, as indicated by the sums. In particular, for the SM gauge bosons ($W, Z$ and transversely polarised 
photon $\gamma$) we have
\begin{equation}
n_W = 6, \qquad n_Z = 3, \qquad n_\gamma = 2 \,,
\end{equation}
whereas for scalars and the longitudinally polarized photon $(A_L)$ we have
\begin{equation}
n_s = 6, \qquad n_{A_L} = 1\,,
\end{equation}
while for fermions
\begin{equation}
n_{u,d,c,s,t,b} = 12, \qquad n_{e,\mu,\tau} = 4\,, \qquad n_{N_{1,\ldots,6}} = 2 \,.
\end{equation}
with $N_{1,\ldots,6}$ denoting the six physical heavy neutrinos in scenario 3.

The presence of $T^2$ terms in the thermal expansion suggests the possibility for symmetry restoration at high temperatures. Furthermore, it typically implies the breakdown of perturbation theory in a close vicinity of the critical temperature. This must be addressed by means of an all-order resummation procedure via the addition of the so called daisy or ring diagrams~\cite{Dolan:1973qd,Parwani:1991gq,Arnold:1992rz,Espinosa:1995se}. In practice, this is done by a correction to the potential mass terms
\begin{equation}
	 \mu_\alpha^2(T) = \mu_\alpha^2 + c_\alpha T^2 \,,
	\label{eq:mu-T}
\end{equation}
where the $c_\alpha$ coefficients can be calculated from \eqref{eq:DV_LO} as follows
\begin{equation}
c_\alpha = \dfrac{\delta^2 {\Delta V^{(1)}(T,\phi_h, \phi_\sigma)}|_{\rm L.O.}}{\delta \phi_\alpha^2}\,.
\end{equation}
While for scenarios 1 and 2 we have
\begin{align} 
& c_h = \frac{3}{16} g^2 + \frac{1}{16} {g'}^2 + 
\frac12 \lambda_\Phi + \frac{1}{12} \lambda_{\Phi \sigma}+ 
\frac14 (y_t^2 + y_b^2 + y_c^2 + y_s^2 + y_u^2 + y_d^2) + \frac{1}{12} (y_{\tau}^2 + y_{\mu}^2 + y_{e}^2)\,, \\
& c_\sigma = \frac13\lambda_\sigma + \frac16 \lambda_{\Phi \sigma}\,,
\label{eq:coeff_thermalmasses}
\end{align}
with $g$ and $g'$ the EW gauge couplings and $y_i$ the Yukawa coupling of the SM particle $i$, for the case of scenario 3 the only relevant modification comes from the neutrino sector where $c_\sigma$ receives an additional contribution from the neutrino Yukawa couplings of the form
\begin{equation} 
c_\sigma \to c_\sigma + \dfrac{1}{24} \sum_{i = 1}^6 Y_{\sigma_i}^2\,.
\label{eq:coeff_thermalmassesII}
\end{equation}
The longitudinal modes of the gauge bosons also receive thermal corrections which look like
\begin{eqnarray}
&& m_{W_L}^2(\phi_{h};T) = m_W^2(\phi_{h}) + \frac{11}{6}g^2T^2\,, \\
&& m_{Z_L,A_L}^2(\phi_{h};T) = \frac{1}{2}m_Z^2(\phi_{h}) + 
\frac{11}{12}(g^2+{g'}^2)T^2 \pm {\cal D} \,,
\end{eqnarray}
with
\begin{equation}
{\cal D}^2 = \Big(\frac{1}{2}m_Z^2(\phi_{h}) + \frac{11}{12}(g^2+{g'}^2)T^2 \Big)^2 - 
\frac{11}{12} g^2{g'}^2 T^2 \Big( \phi_h^2 + \frac{11}{3}T^2 \Big) \,.
\end{equation}

The counterterm Lagrangian $V_{\rm ct}$ is given by
\begin{align}
V_{\rm ct}&=\delta\mu_{\Phi}^{2} \Phi^{\dagger} \Phi+\delta\lambda_{\Phi}\left(\Phi^{\dagger} \Phi\right)^{2}+\delta\mu_{\sigma}^{2} \sigma^{\dagger} \sigma+\delta\lambda_{\sigma}\left(\sigma^{\dagger} \sigma\right)^{2} \nonumber \\
&+\delta\lambda_{\Phi \sigma} \Phi^{\dagger} \Phi \sigma^{\dagger} \sigma+\left(\frac{1}{2} \delta\mu_{b}^{2} \sigma^{2}+\text { h.c. }\right)\, .
\label{eq:countertermpotentialmsbar}
\end{align}
Note that we only perform the renormalization of the potential parameters and leave the fields untouched. The counterterms are fixed by imposing that the Coleman-Weinberg potential and counterterm potential should not change the form of the minimum conditions and masses at zero temperature~\cite{Camargo-Molina:2016moz, Basler:2016obg}
\begin{align}
\left\langle \frac{\partial V_{\rm ct}}{\partial h_i}\right\rangle = \left\langle-\frac{\partial  V^{(1)}_{\rm CW}}{\partial h_i}\right\rangle\,, & & \left\langle \frac{\partial^2 V_{\rm ct}}{\partial h_i\partial h_j}\right\rangle = \left\langle- \frac{\partial^2  V^{(1)}_{\rm CW}}{\partial h_i\partial h_j}\right\rangle\,.
\end{align}
With these conditions, the counterterms for the scalar singlet extension model in the conditions of scenario 1 where $\sigma$ has no VEV at zero temperature, are given by,
\begin{align}
\delta\mu^2_\Phi&=-\frac{3}{2v_h}\frac{\partial V^{(1)}_{\rm CW}}{\partial h} +\frac{1}{2}\frac{\partial^2 V^{(1)}_{\rm CW}}{\partial h^2}\,,  & \delta\lambda_\Phi &= \frac{1}{2v_h^3}\frac{\partial V^{(1)}_{\rm CW}}{\partial h}-\frac{1}{2v_h^2}\frac{\partial^2 V^{(1)}_{\rm CW}}{\partial h^2}\,,\nonumber \\
 \delta\mu_\sigma^2&=0\,, &  \delta\lambda_\sigma&=0\,,\label{eq.ct:scenario1}\\
\delta\lambda_{\Phi\sigma}&=-\frac{2}{v_h^2}\frac{\partial^2 V^{(1)}_{\rm CW}}{\partial \sigma_R^2}\,,  & \delta\mu^2_{b}&=0\,.\nonumber
\end{align}
In the case of a non-zero singlet VEV, i.e. scenario 2 and 3, we get
\begin{align}
\delta\mu^2_\Phi&=-\frac{3}{2v_h}\frac{\partial V^{(1)}_{\rm CW}}{\partial h} +\frac{1}{2}\frac{\partial^2 V^{(1)}_{\rm CW}}{\partial h^2} +\frac{v_\sigma}{2v_h}\frac{\partial^2 V^{(1)}_{\rm CW}}{\partial h\partial \sigma_R}\,,&  \delta\lambda_\Phi &= \frac{1}{2v_h^3}\frac{\partial V^{(1)}_{\rm CW}}{\partial h}-\frac{1}{2v_h^2}\frac{\partial^2 V^{(1)}_{\rm CW}}{\partial h^2} \nonumber\,,\\
 \delta\mu_\sigma^2&=-\frac{3}{2v_\sigma}\frac{\partial V^{(1)}_{\rm CW}}{\partial \sigma_R} +\frac{1}{2}\frac{\partial^2 V^{(1)}_{\rm CW}}{\partial \sigma_R^2} +\frac{v_h}{2v_\sigma}\frac{\partial^2 V^{(1)}_{\rm CW}}{\partial h\partial \sigma_R}\,, &  \delta\lambda_\sigma&= \frac{1}{2v_\sigma^3}\frac{\partial V^{(1)}_{\rm CW}}{\partial \sigma_R}-\frac{1}{2v_\sigma^2}\frac{\partial^2 V^{(1)}_{\rm CW}}{\partial \sigma_R^2}\,, \label{eq.ct:scenario2}\\
 \delta\lambda_{\Phi\sigma}&=-\frac{1}{v_h v_\sigma}\frac{\partial^2 V^{(1)}_{\rm CW}}{\partial h\partial \sigma_R}\,, & \delta\mu^2_{b}&=0\,.\nonumber
\end{align}


There are three temperatures relevant to the phase transition. First the critical temperature at which the effective potential has two degenerate minima. 
Second the nucleation temperature $T_n$. Below the critical temperature the global minimum, that is, the true vacuum emerges and  
the FOPT becomes efficient if the transition probability is of order one per unit Hubble time and Hubble volume.
Hence, at the bubble nucleation temperature, $T_n$, the probability of one transition per cosmological horizon volume is~\cite{Anderson:1991zb},
\begin{eqnarray}
\int_0^{t_n} \Gamma\, V_H(t)\,dt = \int_{T_n}^\infty \frac{dT}{T}
\Big( \frac{2\zeta M_{\rm Pl}}{T} \Big)^4 e^{-\hat{S}_3/T}={\cal O}(1) \,, 
\label{Tn-cond}
\end{eqnarray}
where $V_H(t)$ is the volume of the cosmological horizon, $\zeta\sim 3\cdot 10^{-3}$, 
$M_{\rm Pl}$ is the Planck scale, and
\begin{eqnarray}
\Gamma \sim A(T) e^{-\hat{S}_3/T}\,, \qquad A(T)={\cal O}(T^4) \,,
\end{eqnarray}
is the tunneling rate per unit time per unit volume~\cite{Affleck:1980ac,Linde:1977mm}. The condition (\ref{Tn-cond}) 
numerically translates to the following equation~\cite{Dine:1992wr,Anderson:1991zb,Quiros:1999jp}
\begin{equation}
\frac{\hat{S}_3(T_n)}{T_n} \sim 140 \,,
\label{def_nucleation_temperature}
\end{equation}
which can then be numerically solved with respect to $T_n$.

Finally, another important temperature for the phase transition is the percolation temperature, defined as the temperature at which at least $34\%$ of the false vacuum has tunnelled into the true vacuum~\cite{Ellisupdated} or the probability 
of finding a point that is still in the false vacuum is $70\%$. This condition forces the existence of a large connected structure of true vacuum that spans the whole Universe, at the percolation temperature, such that it 
 cannot collapse back into the false vacuum. This large structure is designated as percolating cluster.
The probability of finding a point in the false vacuum is~\cite{Ellisupdated}
\begin{align}
P(T) = e^{-I(T)}, & & I(T) = \frac{4\pi v_b^3}{3} \int_T^{T_c} \frac{\Gamma(T')dT'}{T'^4 H(T')}\left(\int_T^{T'}\frac{d\tilde{T}}{H(\tilde{T})}\right)^3\,,
\label{eq:Tp}
\end{align}
and therefore, to find the percolation temperature, one has to solve $I(T_*) = 0.34$ or, equivalently, $P(T_*) = 0.7$.

The strength of the phase transition conventionally denoted as $\alpha$, is related to the latent heat released in the FOPT at the bubble percolation temperature $T_*$. It is defined via the trace anomaly~\cite{Hindmarsh:2015qta,Hindmarsh:2017gnf} as follows
\begin{equation}
\alpha = \frac{1}{\rho_\gamma} \Big[ V_i - V_f - \dfrac{T_*}{4} \Big( \frac{\partial V_i}{\partial T} - 
\frac{\partial V_f}{\partial T} \Big) \Big] \,,
\label{alpha}
\end{equation}
where
\begin{equation}
\rho_\gamma = g_* \frac{\pi^2}{30} T_*^4
\end{equation}
is the energy density of the radiation medium at the bubble percolation epoch written as a function of the effective number of relativistic degrees of freedom,~$g_* \simeq 108.75$ for scenario 1 and 2, and $g_* = 114$ for scenario 3 ~\cite{Grojean:2006bp,Leitao:2015fmj,Caprini:2015zlo,Caprini:2019egz}. 
The values of the effective scalar potential before and after the transition takes place, that is, in the symmetric and broken phases, respectively, are written as $V_i\equiv V_{\rm eff}(\phi^i_{h,S};T_*)$ and $V_f\equiv V_{\rm eff}(\phi^f_{h,S};T_*)$. For an in-depth study about the strength of the phase transition and the respective GW signal, see~Ref.~\cite{Ellis:2018mja}.

The second important characteristic of the FOPT is the inverse time-scale of the phase transition denoted as $\beta$ found in units of the Hubble parameter $H$, such that
\begin{equation}
\frac{\beta}{H} = T_*  \left. \frac{\partial}{\partial T} \left( \frac{\hat{S}_3}{T}\right) \right|_{T_*}\,,
\label{betaH}
\end{equation}
where $\hat{S}_3$ is the Euclidean action
\begin{equation}
\hat{S}_3(\hat{\phi},T) = 4 \pi \int_0^\infty \mathrm{d}r \, r^2 \left\{ \frac{1}{2} \left( \frac{\mathrm{d}\hat{\phi}}{\mathrm{d}r} \right)^2 + V_{\rm eff}(\hat{\phi},T) \right\} \,,
\end{equation}
given in terms of a solution of the equation of motion $\hat{\phi}$ which is usually found by calculating the path that minimizes the energy of the field (for more details, see e.g.~Refs.~\cite{Coleman:1977py,Wainwright:2011kj}). Here, $V_{\rm eff}(\hat{\phi},T)$ is the effective potential at a finite temperature $T$ that can be computed for a given particle physics model.

In this work, we consider only the case of non-runaway nucleated bubbles, \textit{i.e.}~infra-luminal wall expansion velocities $v_b < 1$, following the formalism of Ref.~\cite{Caprini:2019egz} in order to estimate the spectrum of primordial GWs. In the considered scenario the intensity of the GW radiation grows with the ratio 
$\Delta v_\phi/T_*$, where
\begin{equation}
    \Delta v_\phi = |v_\phi^f - v_\phi^i|\,, \qquad \phi = h,\sigma
\end{equation}
is the difference between the VEVs of the initial (metastable) and final (stable) phases at the percolation temperature $T_*$. The quantity $\Delta v_\phi/T_*$ 
is another commonly used measure of the strength of the phase transition, particularly relevant for EW baryogenesis. A phase transition is said to be strongly first-order if the order parameter $v_c/T_c > \mathcal{O}(1)$, where $v_c$ is the value of the Higgs VEV calculated at the critical temperature $T_c$. This is the sphaleron suppression criterion that is one of the most important conditions for successful EW baryogenesis. In this work, we consider $\Delta v_\phi/T_*$ as the order parameter instead. This is not only because we have phases with non-zero EW-singlet VEV 
which contribute to the sphaleron suppression but also due to the fact that the actual phase transition starts at $T_n < T_c$, a temperature for which the bubble nucleation rate exceeds that of the 
cosmological expansion, and finishes effectively at $T_\ast < T_n$. Nevertheless, this condition does not necessarily lead to the generation 
of strong and potentially observable GWs. A sizeable GW signal and small $\beta/H$ needs a large bubble wall velocity $v_b$ and a substantial latent heat release which is related to $\alpha$.

In our analysis, we consider only GWs originating from sound shock waves (SW) which are generated by the bubble's violent expansion in the early Universe. According to the discussion in~Ref.~\cite{Caprini:2019egz} their contribution dominates the peak frequency and the peak amplitude in the primordial GW spectrum. Furthermore, bubble wall collision does not give a meaningful contribution to GWs as discussed in Refs.~\cite{Hindmarsh:2017gnf,Ellis:2019oqb} while magnetohydrodynamic turbulence of the early Universe plasma is usually not accounted for due to large theoretical uncertainties~\cite{Caprini:2019egz}.

The primordial GW signals produced in such violent out-of-equilibrium cosmological processes as the FOPTs are redshifted by the cosmological expansion and look today as a cosmic gravitational stochastic background. The corresponding power spectrum~\cite{Grojean:2006bp,Leitao:2015fmj,Caprini:2001nb,Figueroa:2012kw,Hindmarsh:2016lnk}
\begin{equation}
h^2 \Omega_{\rm GW}(f) \equiv \frac{h^2}{\rho_c} \frac{\partial \rho_{\rm GW}}{\partial \log f}\,, 
\end{equation}
where $\rho_c$ is the critical energy density today, can be found for various GW frequencies $f$ by multiplying the peak amplitude $h^2 \Omega_\mathrm{GW}^\mathrm{peak}$ by the spectral function and reads
\begin{equation}
	h^2 \Omega_\mathrm{GW} = h^2 \Omega_\mathrm{GW}^\mathrm{peak} \left(\dfrac{4}{7}\right)^{-\tfrac{7}{2}} \left(\dfrac{f}{f_\mathrm{peak}}\right)^3 \left[1 + \dfrac{3}{4} \left(\dfrac{f}{f_\mathrm{peak}}\right) \right]^{-\tfrac{7}{2}}\,,
	\label{eq:spectrum}
\end{equation}
where $f_\mathrm{peak}$ is the peak-frequency. Semi-analytic expressions 
for peak-amplitude and peak-frequency in terms of $\beta/H$ and $\alpha$ can be found in Ref.~\cite{Caprini:2019egz} and can be summarised as follows
\begin{eqnarray}
	&& f_\mathrm{peak} = 26 \times 10^{-6} \left( \dfrac{1}{H R} \right) \left( \dfrac{T_\mathrm{n}}{100} \right) \left( \dfrac{g_\ast}{100~\mathrm{GeV}} \right)^{\tfrac{1}{6}} \mathrm{Hz} \,, \label{eq:fpeak} \,, \\
    && h^2 \Omega_\mathrm{GW}^\mathrm{peak} = 1.159 \times 10^{-7} \left(\dfrac{100}{g_\ast}\right)  \left(\dfrac{HR}{\sqrt{c_s}}\right)^2 K^{\tfrac{3}{2}} \qquad \rm{for} \qquad H \tau_\mathrm{sh} = \dfrac{2}{\sqrt{3}} \dfrac{HR}{K^{1/2}} < 1 \,,
    \\
    && h^2 \Omega_\mathrm{GW}^\mathrm{peak} = 1.159 \times 10^{-7} \left(\dfrac{100}{g_\ast}\right)  \left(\dfrac{HR}{c_s}\right)^2 K^{2} \qquad \rm{for} \qquad H \tau_\mathrm{sh} = \dfrac{2}{\sqrt{3}} \dfrac{HR}{K^{1/2}} \simeq 1 \,,
\label{eq:Opeak2}
\end{eqnarray}
where $\tau_\mathrm{sh}$ is the fluid turnover time or the shock formation time, which quantifies the time the GW source was active. In these expressions, $c_s = 1/\sqrt{3}$ is the speed of sound, $R$ is the mean bubble separation,
\begin{equation}
	K = \dfrac{\kappa \alpha}{1 + \alpha}
	\label{eq:K}
\end{equation}
is the fraction of the kinetic energy in the fluid to the total bubble energy, and
\begin{equation}
	H R = \dfrac{H}{\beta} \left( 8 \pi \right)^{\tfrac{1}{3}} \max\left(v_b, c_s\right) \,.
	\label{eq:HR}
\end{equation}
where $\kappa$ is the efficiency factor that can be found in Ref.~\cite{Espinosa:2010hh}.

The bubble wall velocity has to be rather large to give rise to detectable GWs spectra although it is quite challenging to provide a precise estimate for it~\cite{Dorsch:2018pat,Moore:1995ua}. Our analysis is performed using \texttt{CosmoTransitions}~\cite{Wainwright:2011kj}, 
considering the case of supersonic detonations in order to maximize the GW peak amplitude where the wall velocity $v_b$ is taken to be above the Chapman-Jouguet limit,
\begin{equation}
	v_\mathrm{J} = \dfrac{1}{1+\alpha} \left(c_s + \sqrt{\alpha^2 + \tfrac{2}{3} \alpha}\right)\,.
	\label{eq:vJ}
\end{equation}
For certain parameter configurations, one also expects sequential phase transition patterns potentially leading to multi-peak GWs spectra~\cite{Vieu:2018zze,Morais:2019fnm,Greljo:2019xan,Aoki:2021oez}.

\section{Analysis of Scenario 2}

The first goal of this study is to understand in which of the three scenarios a strong FOPT leading to the detections of gravitational waves
in future experiments is realisable. We have concluded that only in scenarios 2 and 3 do one finds GW that can be probed by those experiments. Since scenario
3 was already discussed in~\cite{Addazi:2019dqt}  we will now focus on scenario 2.
We will come back to scenario 3 when discussing the variation of the GW peak and frequency with the SM parameters.

\subsection{GWs detection}
\label{sec:GWSc1}

In Table~\ref{tab2} we show the ranges of the input parameters in the scans for scenario 2. The range for $\mu_{b}^2$ reflects a variation
of the mass of the DM candidate between 50 GeV and 1 TeV. The range of variation of $\theta$ takes into account the LHC Higgs couplings measurements
that force the 125 GeV Higgs, dubbed as $h_1$, to be very SM-like with $\cos \theta > 0.85$. 
\begin{table*}[h!]
\centering
\begin{tabular}{@{}rcccr@{}}\toprule
& \multicolumn{3}{c}{Scenario 2} \\
\cmidrule{2-4} 
& Parameter & Range & Distribution &\\ \midrule
& $m_{h_2}$ & $[50,\,1000]\,\text{GeV}$ & linear\\
& $v_{\sigma}$ & $[50,\,1000]\,\text{GeV}$ & linear &\\
& $\mu_{b}^2$ & $[-500000,\,-1250]\,\text{GeV}^2$ & linear &\\
& $\theta$ & $[-\arccos{(0.85)},\,	\arccos{(0.85)}]$ & linear &\\
\bottomrule
\end{tabular}
\caption{Ranges of the input parameters in the scans for scenario 2.}
\label{tab2}
\end{table*}
We will also show in the plots the reach predicted by LISA, by the Deci-Hertz Interferometer Gravitational Wave Observatory (DECIGO)~\cite{Kawamura:2011zz} 
and by the Big Bang Observer (BBO)~\cite{BBO} proposed as a follow-on mission to LISA. There is also a more recent proposal under discussion,
the TianQin Observatory~\cite{Liang:2021bde}.

Having set the stage for the calculation of the GW power spectrum we will now examine which parameters play an important role in the detection of GWs in the near future.
Let us start by noting that from all the possible 16 phase transition patterns we found that in the considered scenario
the ones that are above the LISA line are 148 of the type $(0,0) \to (0,v_\sigma^f)$, 29  $(v_h^i,0) \to (v_h^f, v_\sigma^f)$, 14 $(0,v_\sigma^i) \to (v_h^f, v_\sigma^f)$, 
11 $(0,v_\sigma^i) \to (0, v_\sigma^f)$, 2 $(0,v_\sigma^i) \to (v_h^f, v_\sigma^i)$, 1 $(0, 0) \to (v_h^f, 0)$ and 1 of the type $(v_h^i, 0) \to (v_h^i, v_\sigma^f)$, where
the superscript $i$ and $f$ mean initial and final~\footnote{The first pair of values represent the VEVs before the phase transition and second pair are the values after the phase transition. The first term
in the pair is the doublet VEV while the second is the real part of the singlet VEV (the imaginary component of the singlet has always zero VEV). }.
Clearly, most transitions prefer a final phase with a non-vanishing singlet VEV.

Before proceeding to the presentation of the results, we need to discuss the calculation of the inverse time-scale of the phase transition in units of the Hubble parameter $H$, $\beta/H$. 
The value of $\beta/H$ is obtained from Eq.~(\ref{betaH}) and it is calculated numerically, resorting to the values of the action provided by the  \texttt{CosmoTransitions} code. 
As can be seen from the expression, the calculation involves the derivative of the action, which means that $\hat{S}_3$ has to be a continuous function in the vicinity of the percolation temperature.
This is not the case -- the action obtained from \texttt{CosmoTransitions} is irregular and we have devised a method to smoothen the action before performing the derivative. This 
procedure is described in detail in appendix~\ref{app:act} together with the estimation of the error in the calculation of  $\beta/H$. All results presented, unless otherwise stated, exclude points
with an error above 25\% (see appendix~\ref{app:act} for further details).

\begin{figure}[h!]
\centering
\includegraphics[width = 0.44 \linewidth]{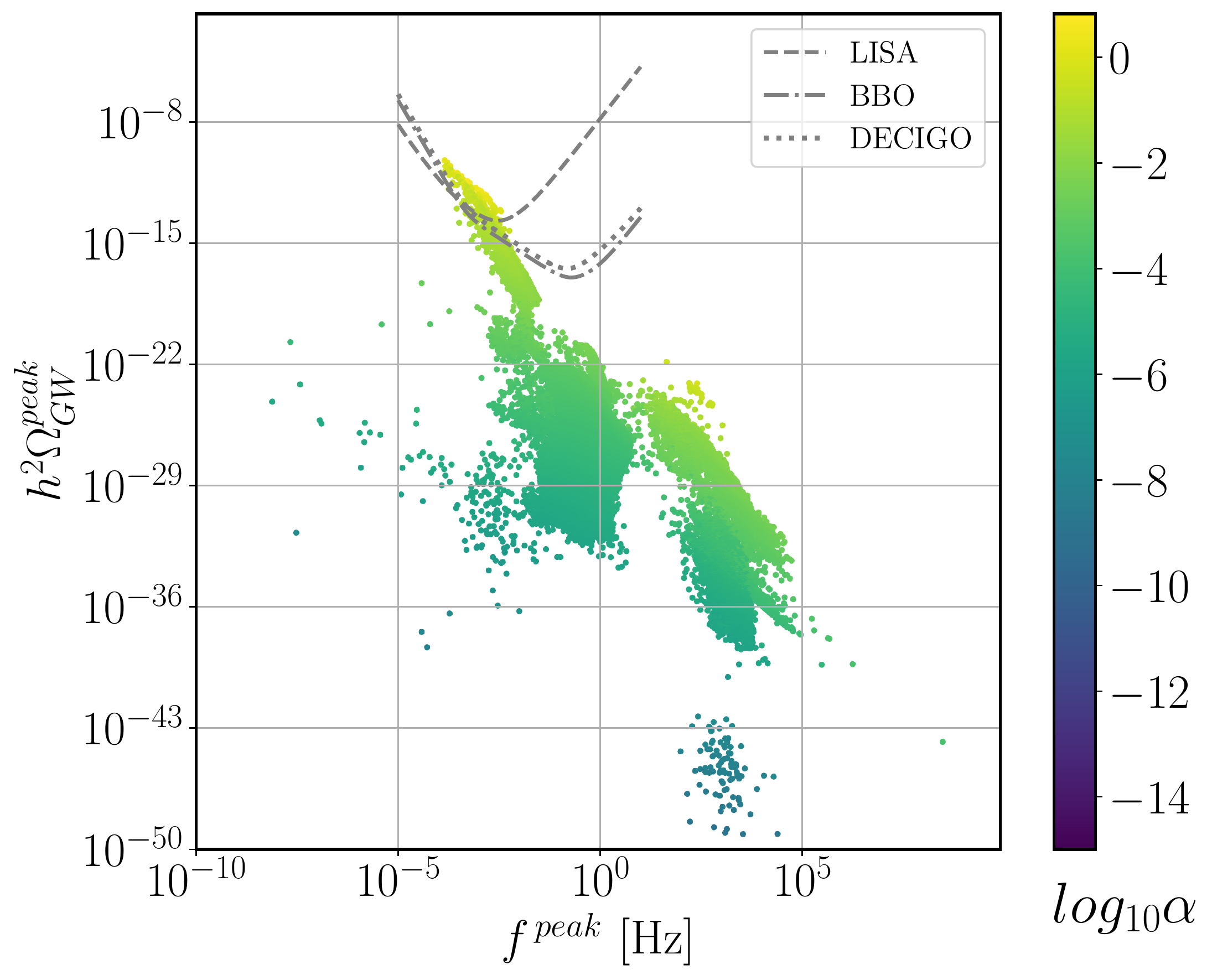}
\includegraphics[width = 0.465 \linewidth]{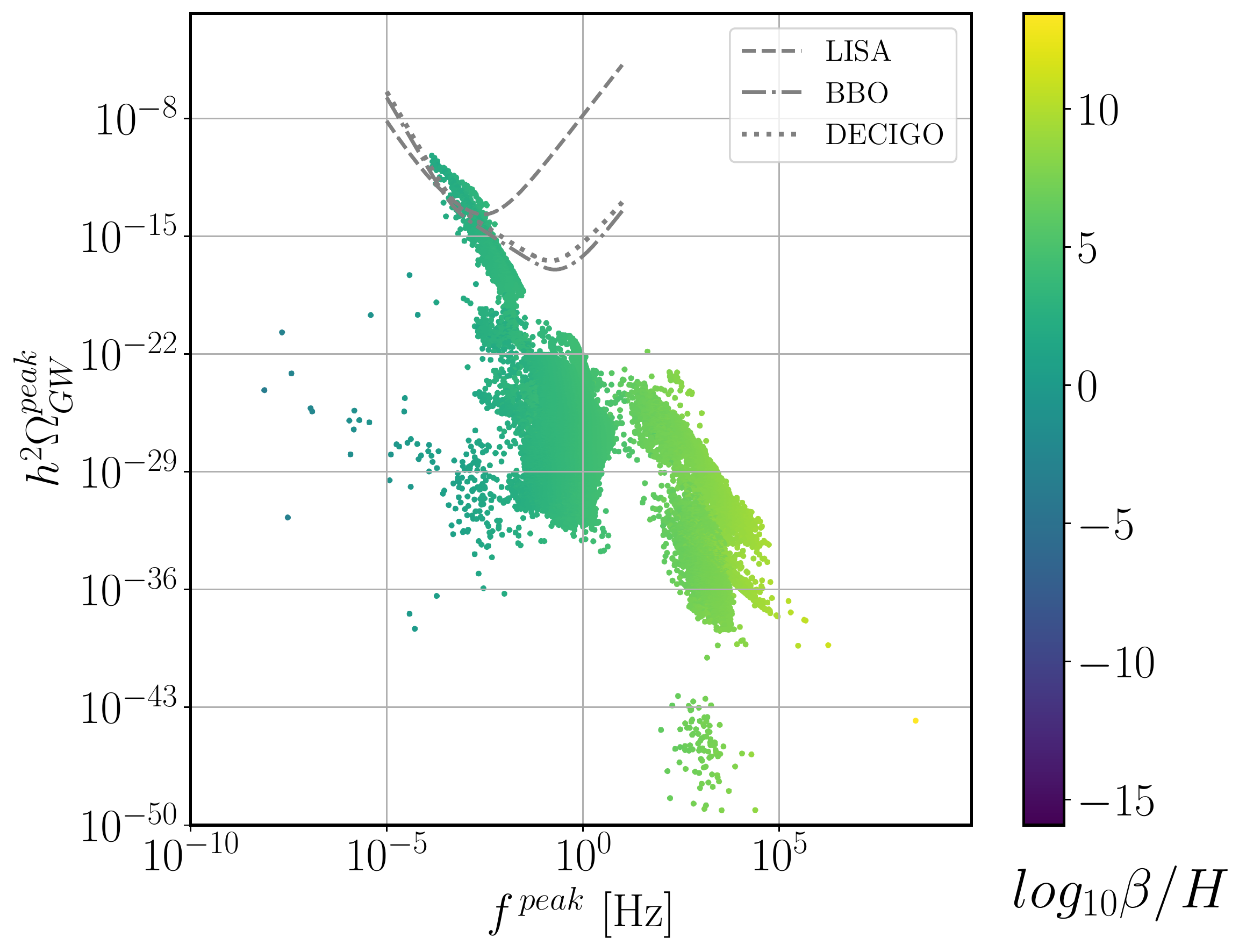}
\caption{The peak-amplitude of the GW signal $h^2 \Omega^{\rm peak}_{\rm GW}$ as a function of the peak frequency $f_{\rm peak}$ in logarithmic scale.
The colour bar indicates the strength of the phase transition $\alpha$ (left panel) and the inverse time-scale of the phase transition in units 
of the Hubble parameter $H$, $\beta/H$ (right panel). The PISCs for LISA BBO and DECIGO are represented with dashed, dot-dashed and dotted lines respectively.}
\label{Figalfbet}
\end{figure}

In Fig.~\ref{Figalfbet} we present the GW signal $h^2 \Omega^{\rm peak}_{\rm GW}$ as a function of the peak frequency $f_{\rm peak}$ in logarithmic scale. The colour bar shown in the scatter plots represents the strength of the phase transition $\alpha$ (left panel) and the inverse time-scale of the phase transition in units of the Hubble parameter $H$, $\beta/H$ (right panel). From the figure it is clear that only values of $\alpha$ above about $0.1$ may lead to GW signals detectable in the near future~\footnote{To be precise, we found points within LISA reach in the range $0.1\leq \alpha \leq 6.6$.}. As for the inverse time-scale, the points within LISA reach are in the range  $34 \leq \beta/H \leq 2257$. The grey curves in these and all remaining plots represent the \textit{peak integrated sensitivity curves} (PISCs) for sound waves recently derived in~Ref.~\cite{Schmitz:2020syl}.

\begin{figure}[h!]
\centering
\includegraphics[width = 0.45 \linewidth]{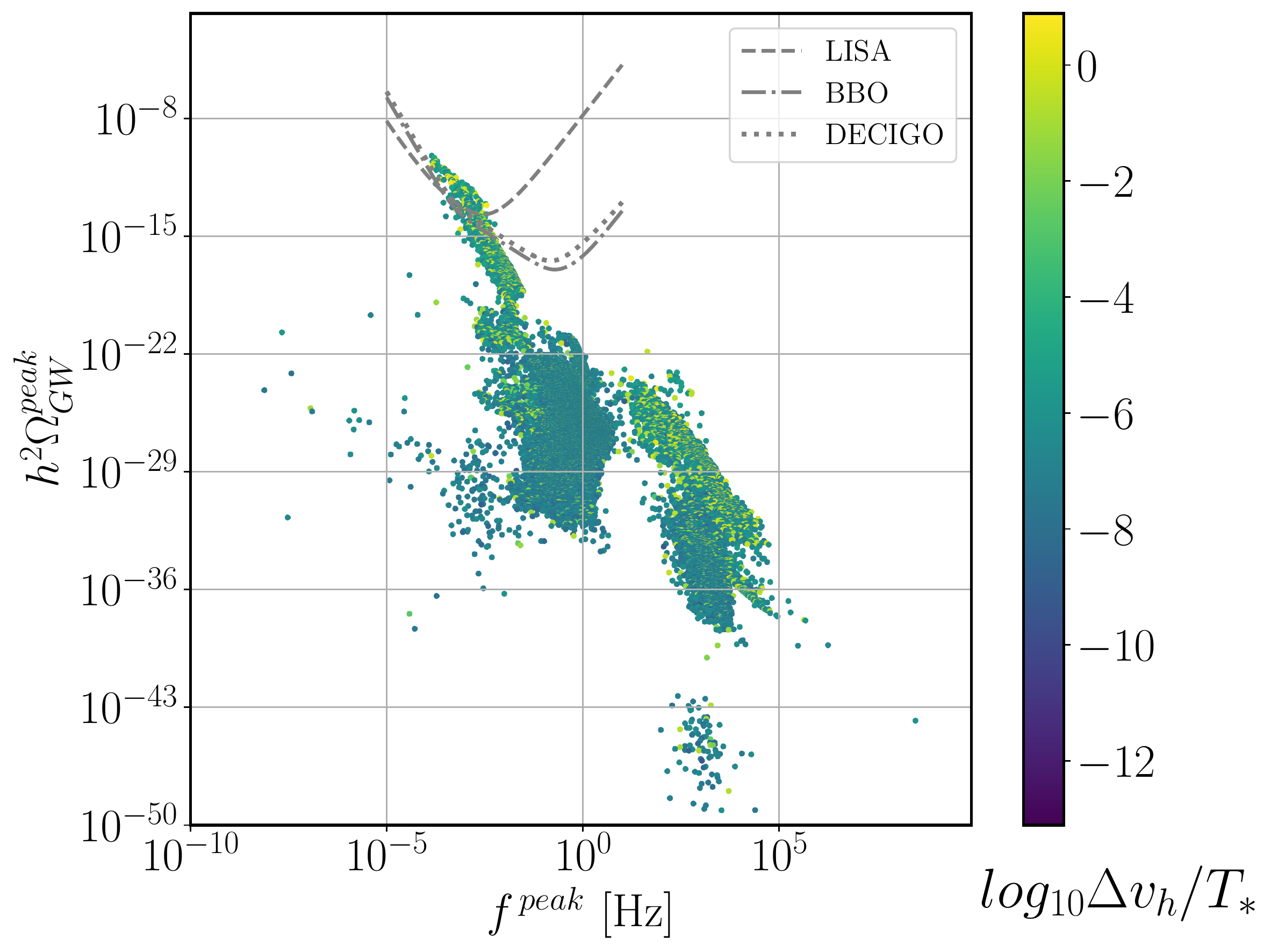}
\includegraphics[width = 0.45 \linewidth]{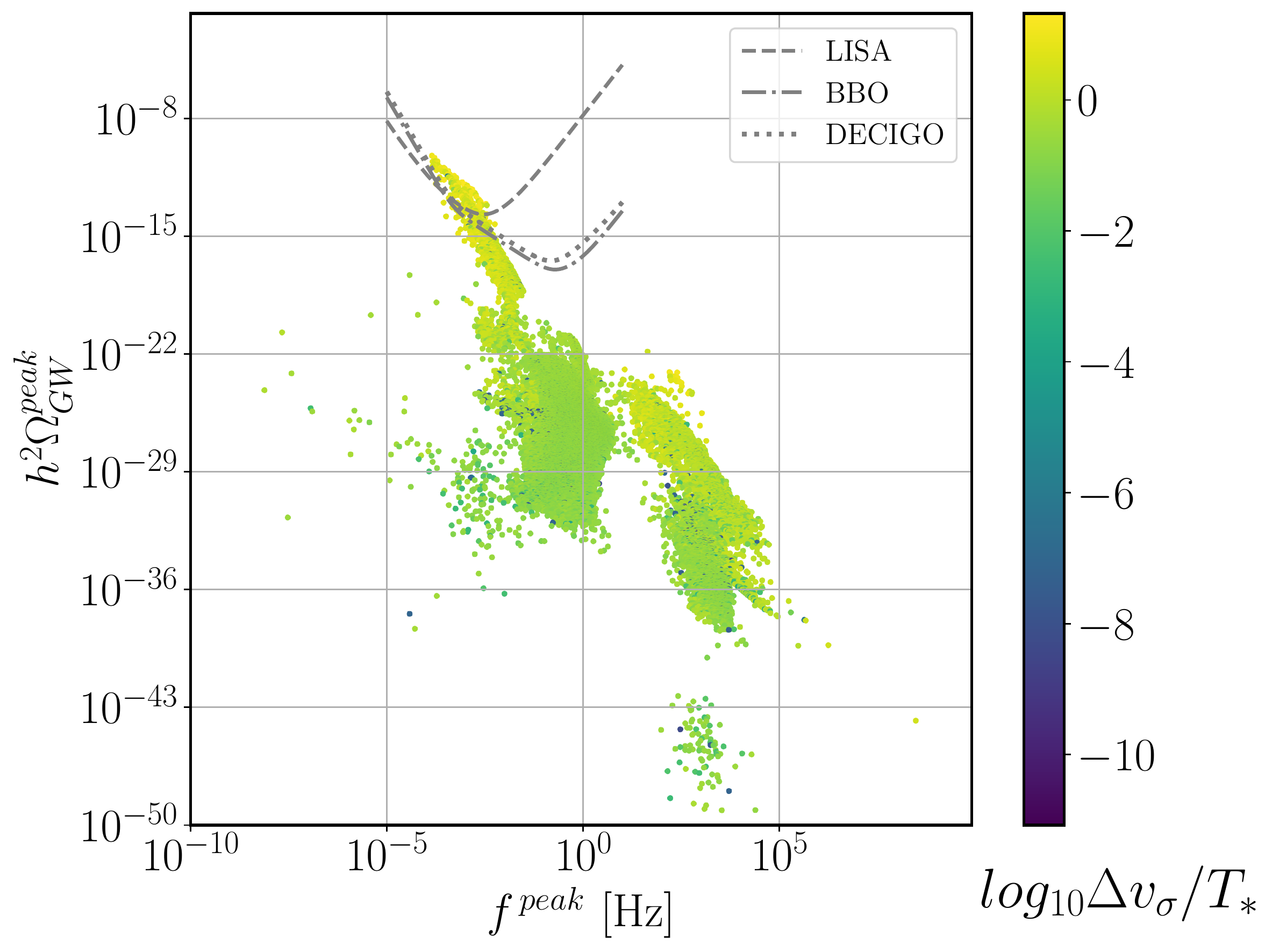}
\caption{The peak-amplitude of the GW signal $h^2 \Omega^{\rm peak}_{\rm GW}$ as a function of the peak frequency $f_{\rm peak}$ in logarithmic scale.
The scatter plots present, in the colour bar, $\Delta v_h/T_*$  (left panel) and $\Delta v_\sigma/T_*$  (right panel). }
\label{Figdeltav}
\end{figure}

\begin{figure}[h!]
\centering
\includegraphics[width = 0.45 \linewidth]{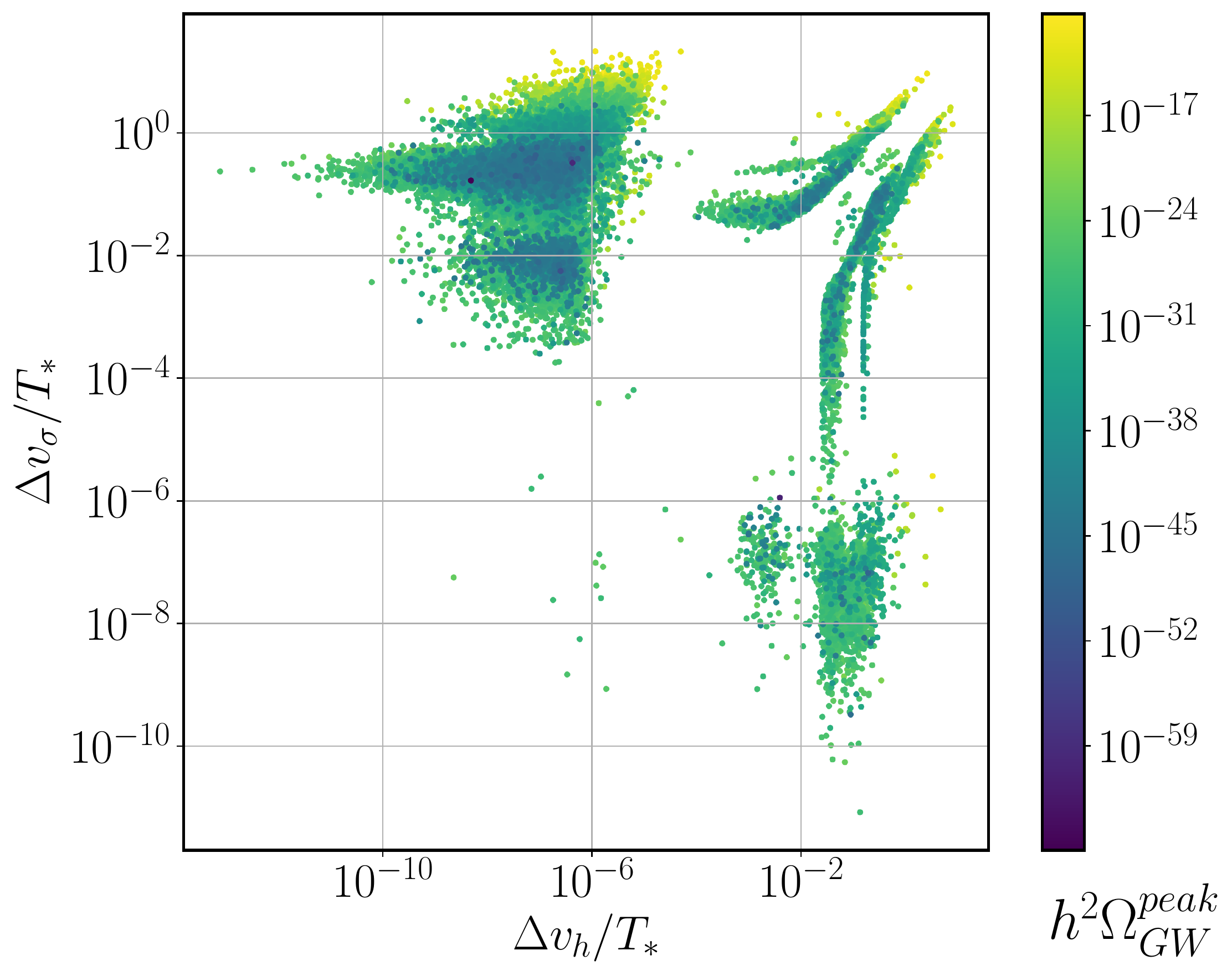}
\caption{Scatter plot showing $\Delta v_h/T_*$  vs. $\Delta v_\sigma/T_*$  with the strength of the GW signal 
$h^2 \Omega^{\rm peak}_{\rm GW}$ given by the colour bar.}
\label{Figdeltadelta}
\end{figure}

In Fig.~\ref{Figdeltav} we again present the GW signal $h^2 \Omega^{\rm peak}_{\rm GW}$ as a function of the peak frequency $f_{\rm peak}$ in logarithmic scale but now the colour scale represents $\Delta v_h/T_*$  (left panel) and $\Delta v_\sigma/T_*$  (right panel). 
As expected the points within reach have large values of $\Delta v_h/T_*$ and of  $\Delta v_\sigma/T_*$ . A clearer picture is obtained when we plot these two variables in the same plot. This is done in Fig.~\ref{Figdeltadelta}
and it shows that it is enough to have a large variation for the singlet, $\Delta v_\sigma/T_*$,  to generate  observable GW signals.

Using our polynomial interpolation of the action and plugging it into \eqref{eq:Tp}, we have observed that the points above LISA have a difference that is below 7 GeV and can even be close to zero in some cases.
%
\begin{figure}[h!]
\centering
\includegraphics[width = 0.45 \linewidth]{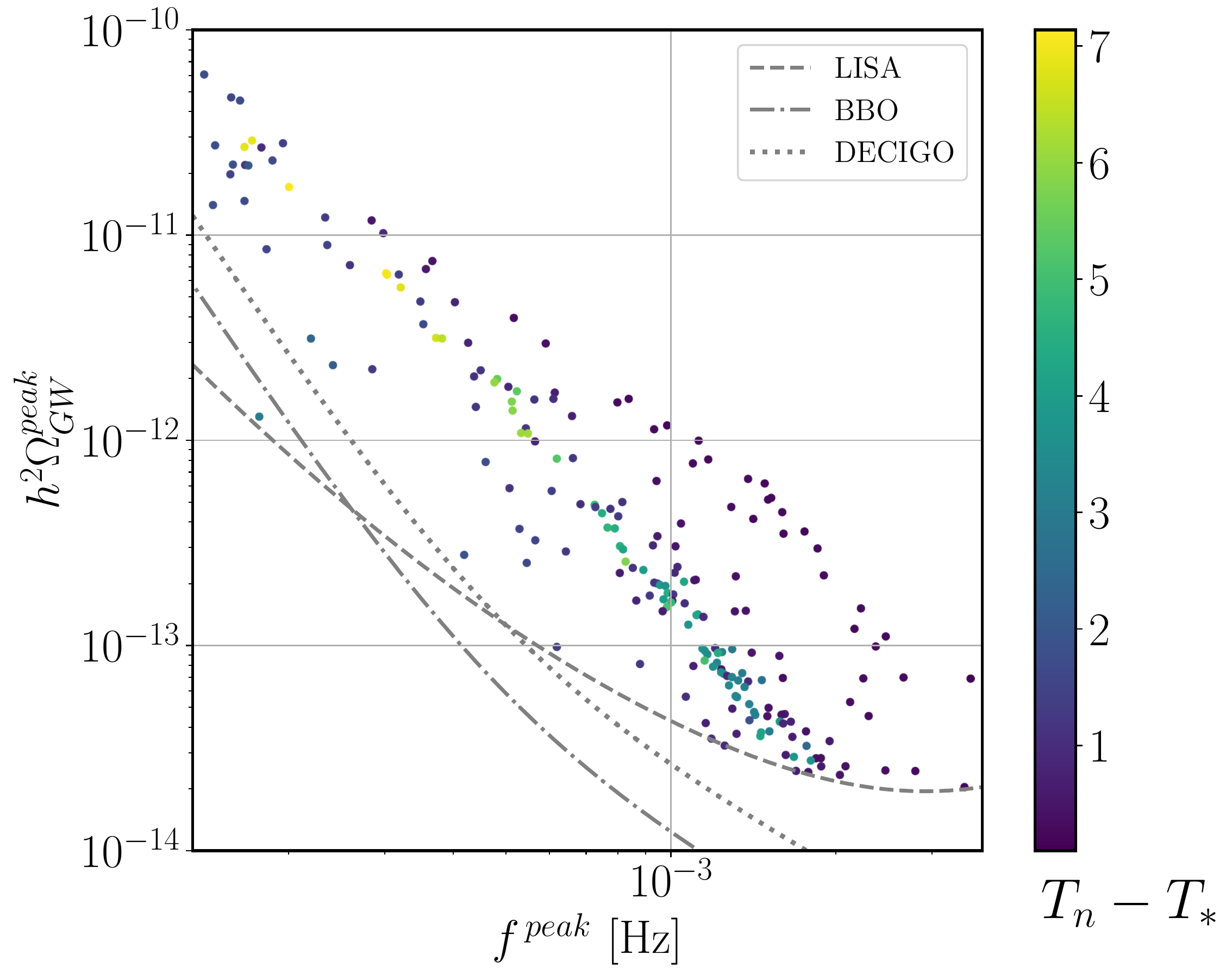}
\caption{Peak-amplitude of the GW signal $h^2 \Omega^{\rm peak}_{\rm GW}$ as a function of the peak frequency $f_{\rm peak}$ in logarithmic scale
with $T_n - T_*$  in the colour bar and given in GeV.}
\label{Figa}
\end{figure}
%
This can be seen in Fig.~\ref{Figa} where we display the peak-amplitude of the GW signal as a function of the peak frequency with $T_n - T_*$  in the colour bar. It is clear that
there is no special trend regarding the difference between the two temperatures in relation to the peak except that above a certain temperature difference there are no points above LISA in our numerical simulations. Note that some of our points, in particular those with larger temperature differences, tend to agree with the calssification in~ \cite{Wang:2020jrd} with respect to supercooling scenarios where we find $\alpha \gtrsim 0.1$ within the LISA sensitivity range.

\subsection{The dark sector}
\label{sec:dark}

\begin{figure}[h!]
\centering
\includegraphics[width = 0.45 \linewidth]{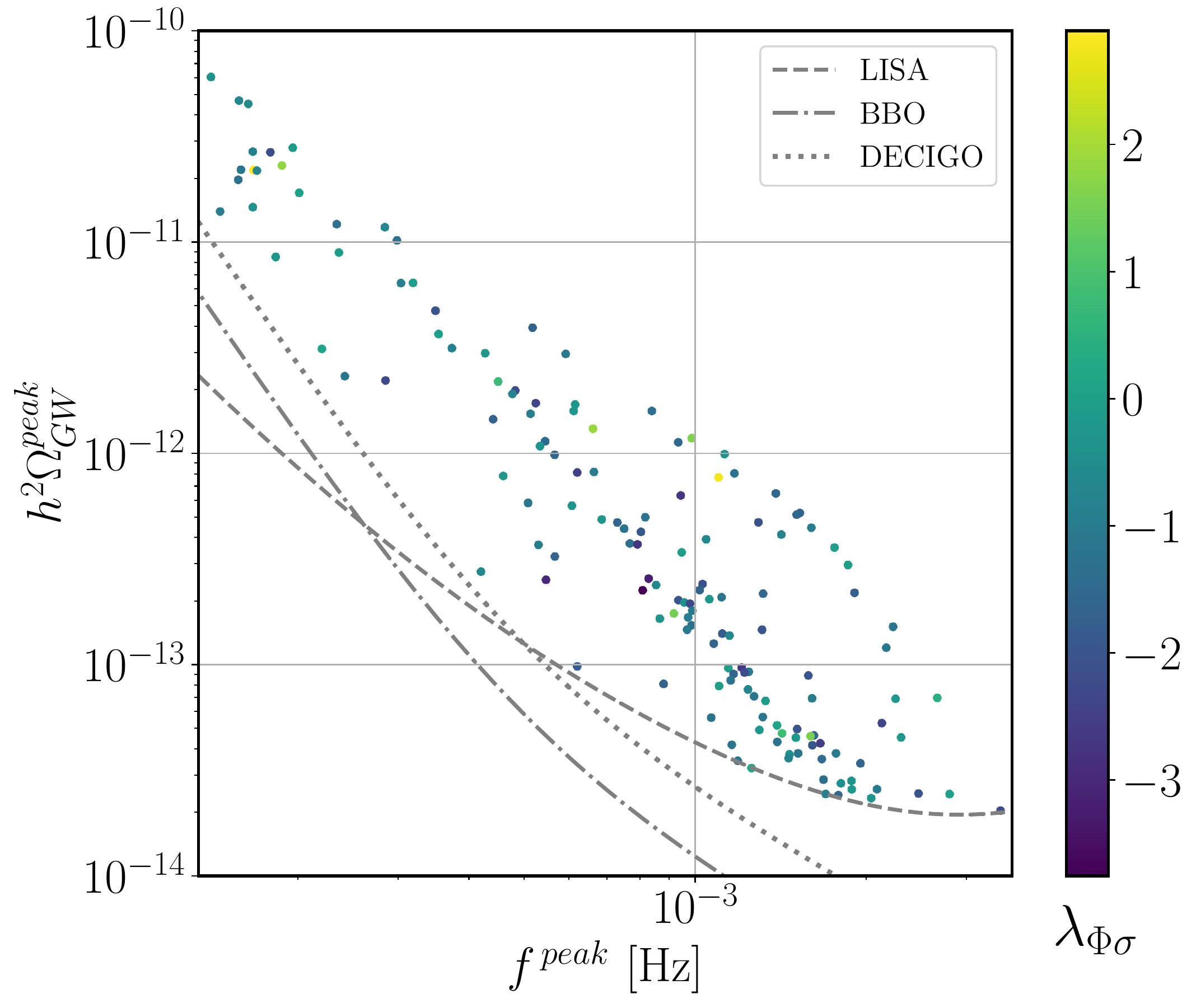}
\includegraphics[width = 0.48 \linewidth]{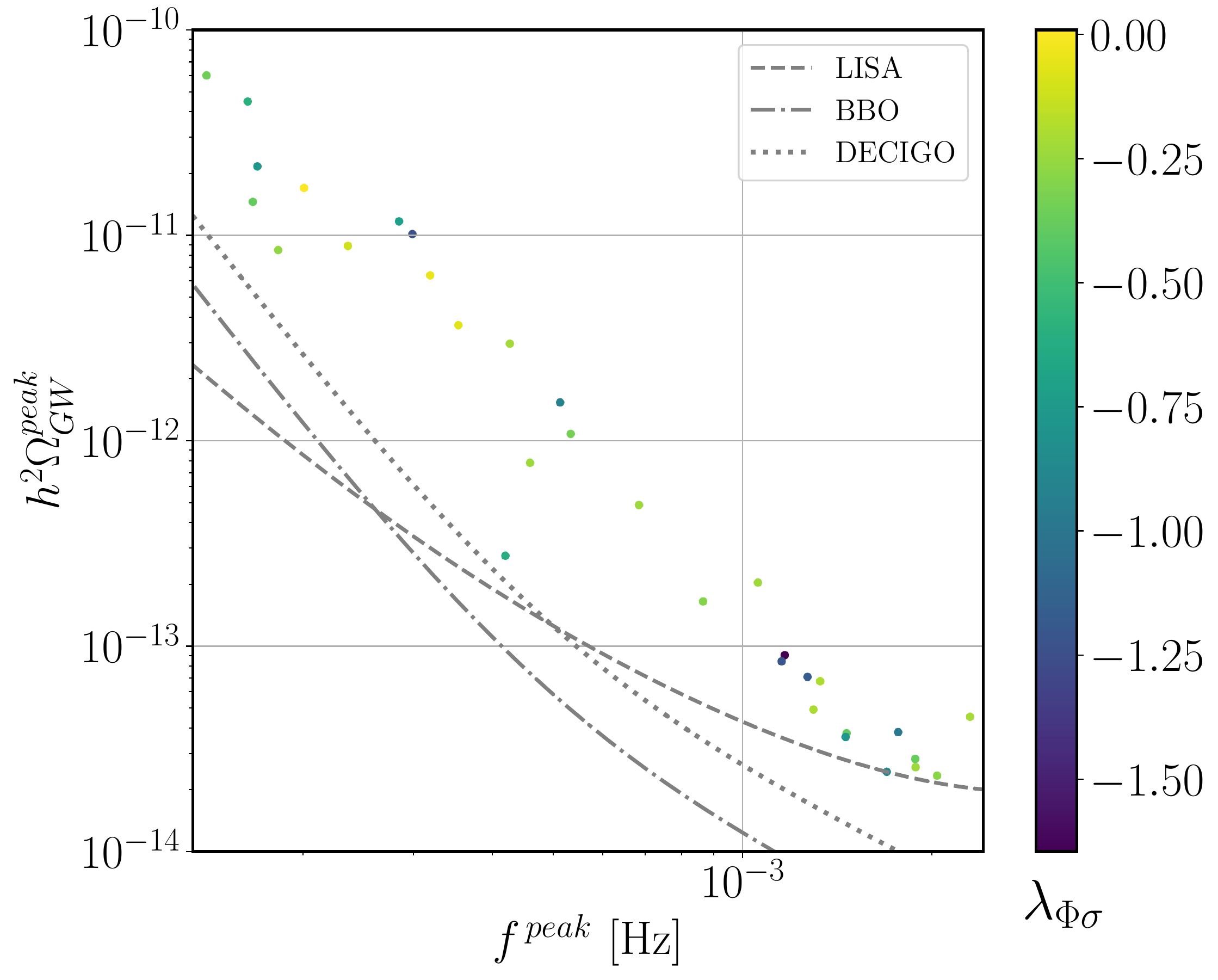}
\caption{
Peak-amplitude of the GW signal $h^2 \Omega^{\rm peak}_{\rm GW}$ as a function of the peak frequency $f_{\rm peak}$ in logarithmic scale
with the portal coupling $\lambda_{\Phi \sigma}$ in the color bar. In the left plot we have set $\lambda_\sigma<1$ and in the right plots the point obey $\lambda_\Phi<1$.}
\label{Figb}
\end{figure}

Let us now discuss the impact of the parameters of the dark sector on possible detection of GWs originating from a FOPT. 
In Fig.~\ref{Figb} we present peak-amplitude of the GW signal as a function of the peak frequency 
with the portal coupling $\lambda_{\Phi \sigma}$ in the color bar. In the left plot we have set $\lambda_\sigma<1$ and in the right plots the points obey $\lambda_\Phi<1$.
We see that there is no trend but it is clear that large values for the peak are obtained with all quartic couplings sufficiently small.
%
%
Although we could not find in our scan points with all quartic couplings below 1, points with two quartic couplings below 1 and the third one
below 2 were common. This is important because as shown in~\cite{Costa:2014qga} if all couplings are below 1 the model is stable up to Planck scale
 and if they are all below 2 the model is stable to slightly below or in the GUT scale. The stability study~\cite{Costa:2014qga}  was 
 performed for the complex singlet extension of the SM using the full two-loop renormalization group equations. Therefore we believe the values 
 shown for quartic couplings should be stable up to somewhere between the GUT and the Planck scale. Moreover, it is possible
 that with  more time of running we would find points with all quartic couplings below 1.

\begin{figure}[h!]
\centering
\includegraphics[width = 0.45 \linewidth]{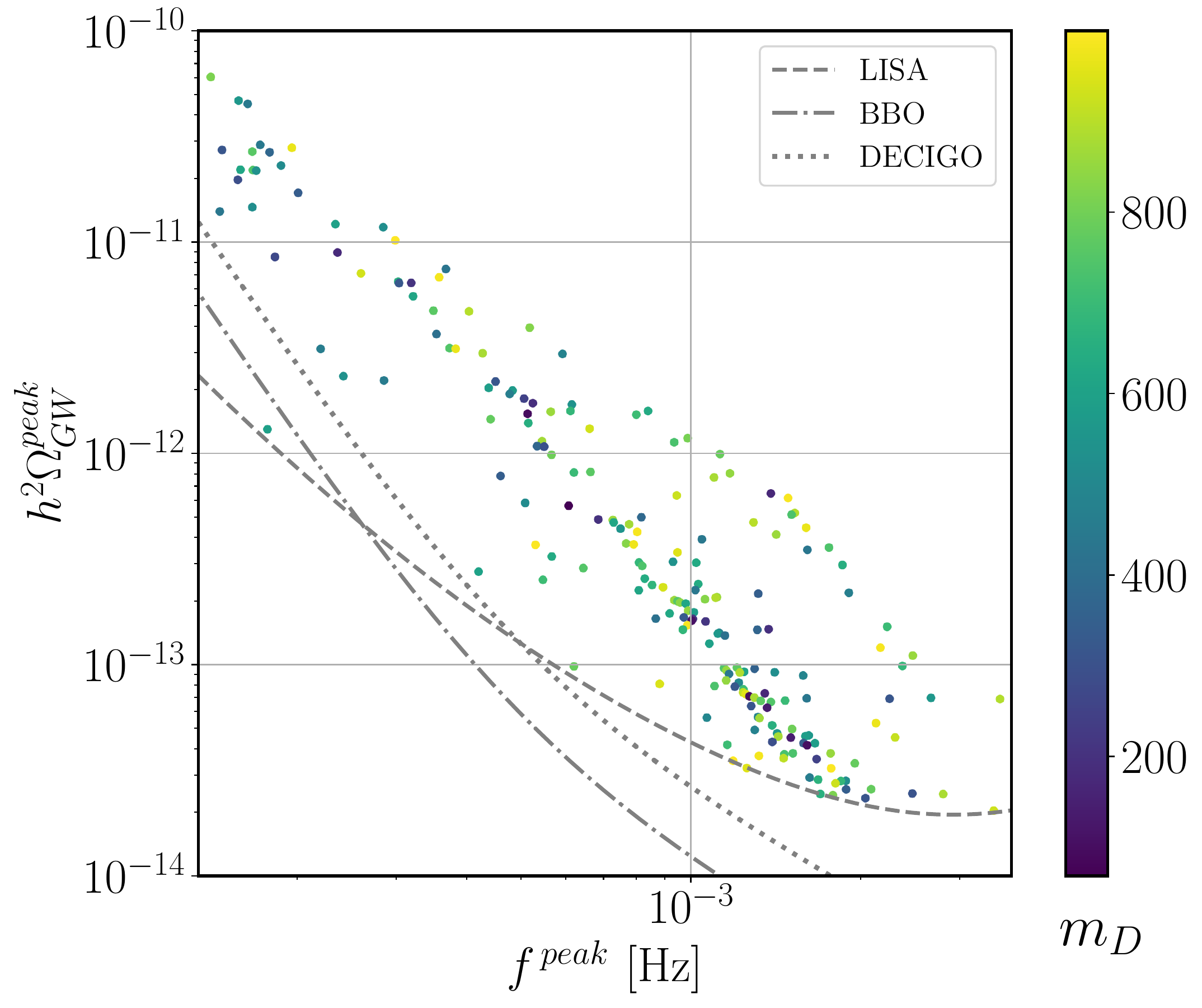}
\includegraphics[width = 0.45 \linewidth]{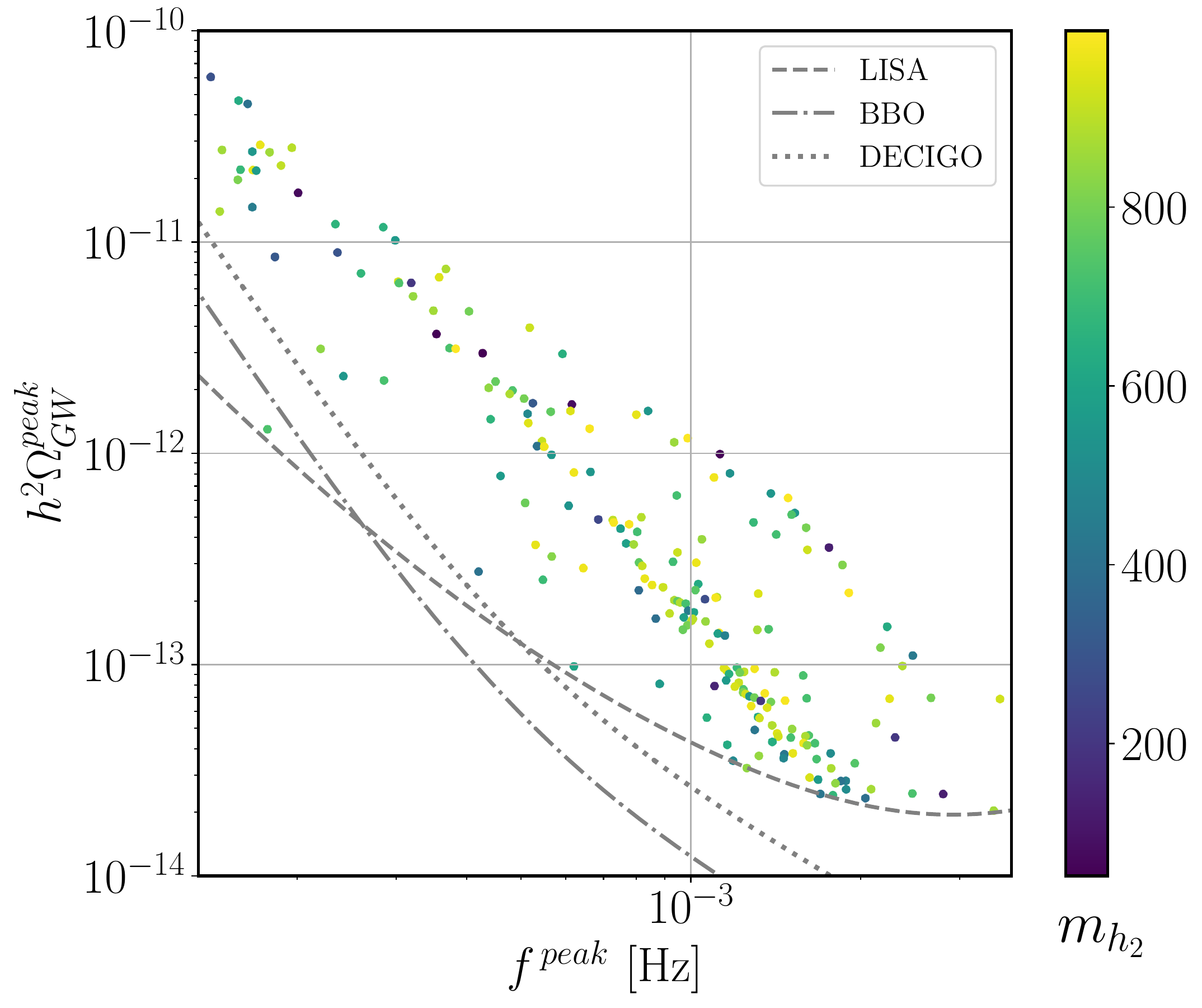}
\caption{The peak-amplitude of the GW signal $h^2 \Omega^{\rm peak}_GW$ as a function of the peak frequency $f_{\rm peak}$ in logarithmic scale.
The scatter plots present, in the colour bar, the dark matter mass $m_D$  (left panel) and non-SM Higgs $m_{h_2}$  (right panel). Masses are expressed in GeV.}
\label{Figc}
\end{figure}

In Fig.~\ref{Figc} we present the peak-amplitude of the GW signal as a function of the peak frequency in logarithmic scale.
The left panel of the scatter plot shows the behaviour with the DM mass, $m_{D}$, while the right panel
refers to the second Higgs mass $m_{h_2}$. Remember that the range of variation chosen for both masses is between 50 GeV and 1 TeV.
The figure shows no particularly interesting pattern with the values for the parameters within the initial chosen range. The allowed values
of the parameters for a strong FOPT with points above LISA are $52.0 \text{ GeV} < m_{h_2} < 997.2 \text{ GeV}$ and $68.2 \text{ GeV} < m_{D} < 999.8 \text{ GeV}$.

\begin{figure}[h!]
\centering
\includegraphics[width = 0.48 \linewidth]{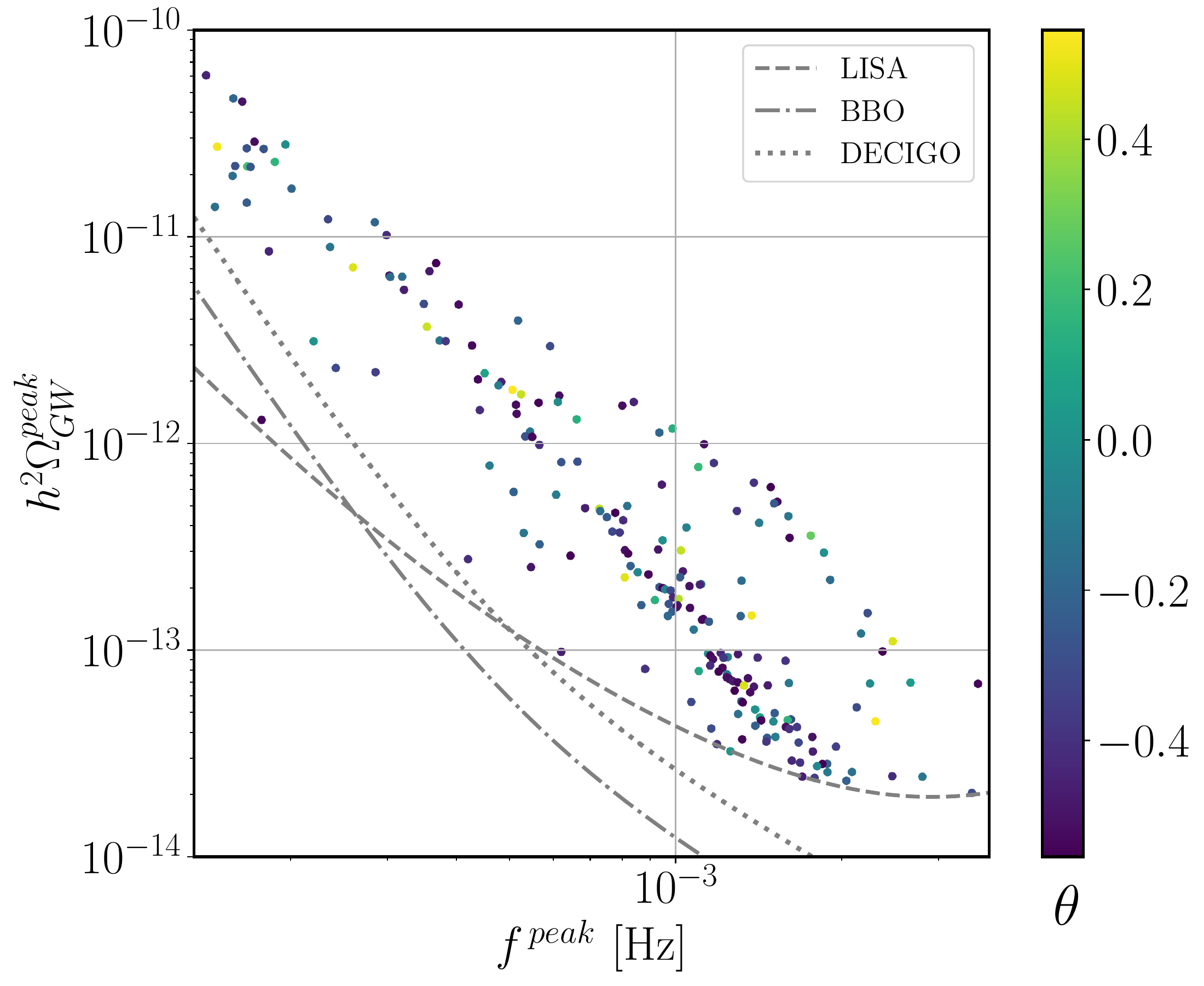}
\includegraphics[width = 0.46 \linewidth]{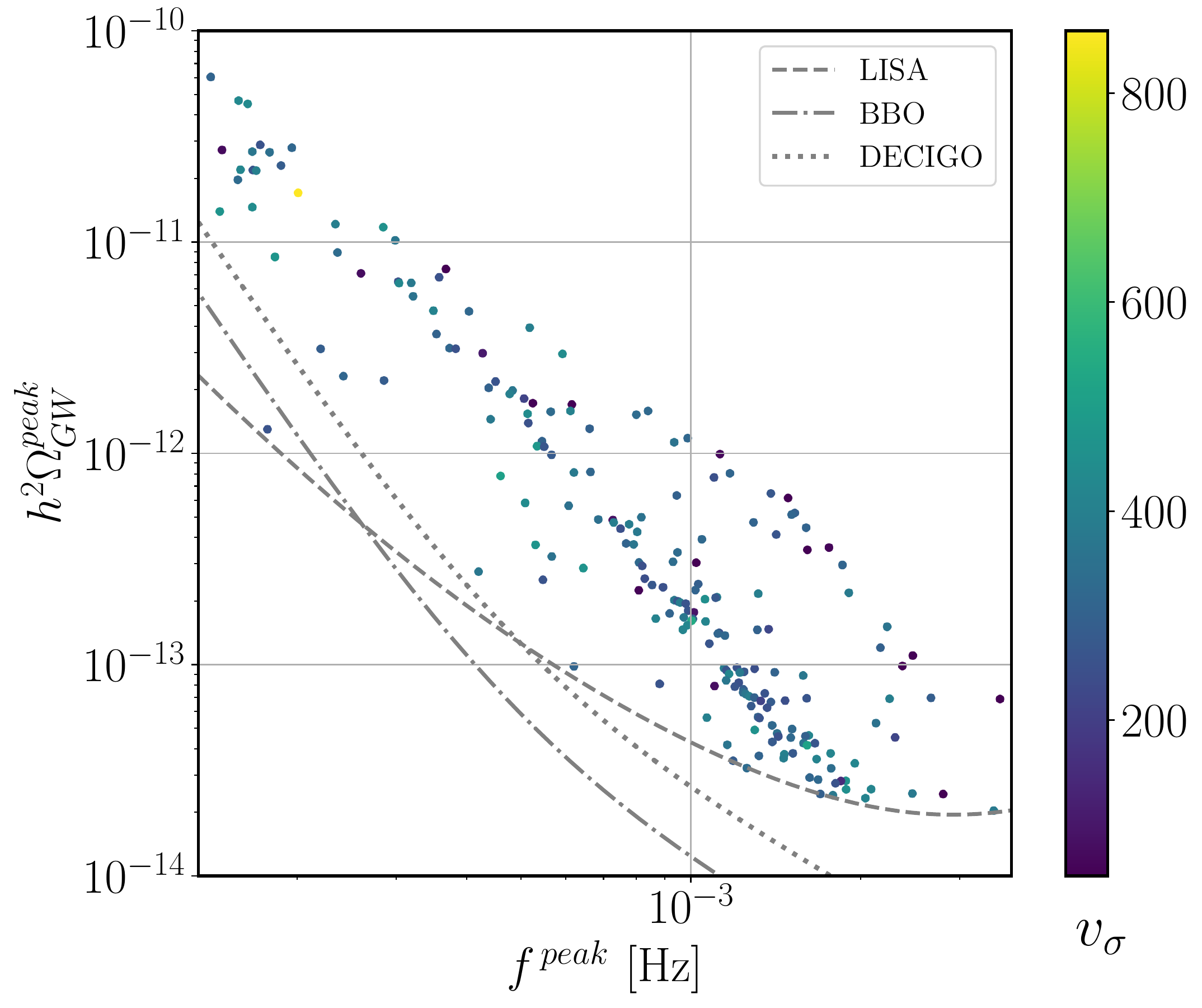}
\caption{Same as Fig.~\ref{Figc} but now as a function of the mixing angle $\theta$ (left) and as a function of $v_\sigma$ (right). }
\label{Figd}
\end{figure}

In Fig.~\ref{Figd} we show the same set of points but now as a function of the mixing angle $\theta$ (left) and as a function of $v_\sigma$ (right).
Again no particular pattern emerges although a slight preference for negative values of theta exists. The values of $v_\sigma$
that allow for a strong FOPF are in the range  $50.8 \text{ GeV} < v_\sigma < 860.1 \text{ GeV}$.

\section{The dependence on the SM parameters}
\label{sec:SMpar}

\subsection{Scenario 2}

In the studies presented in the literature connecting FOPT with the detection of primordial GWs, the role of the SM parameters is never discussed (to the best of our knowledge).
So suppose that a GW is detected and that it points to a given class of models. One should then ask: what happens if we vary each of the SM parameters within the experimentally determined error? Will it lead to significant changes in the characteristics of the GW or is it negligible? These are the questions we will answer in this section taking as benchmark scenario 2. Note that this scenario is not only one of the simplest extensions of the SM but it also features couplings of the Higgs boson to the SM particles that are 
all modified by a common factor $\cos \theta$ that is getting closer and closer to unity. Moreover, as we have seen in the previous section, this is a model that leads to GW signals potentially detectable in not-so-distant future. 
Scenario 3 will be the subject of the next section. 
  
\begin{figure}[h!]
\centering
\includegraphics[width = 0.80 \linewidth]{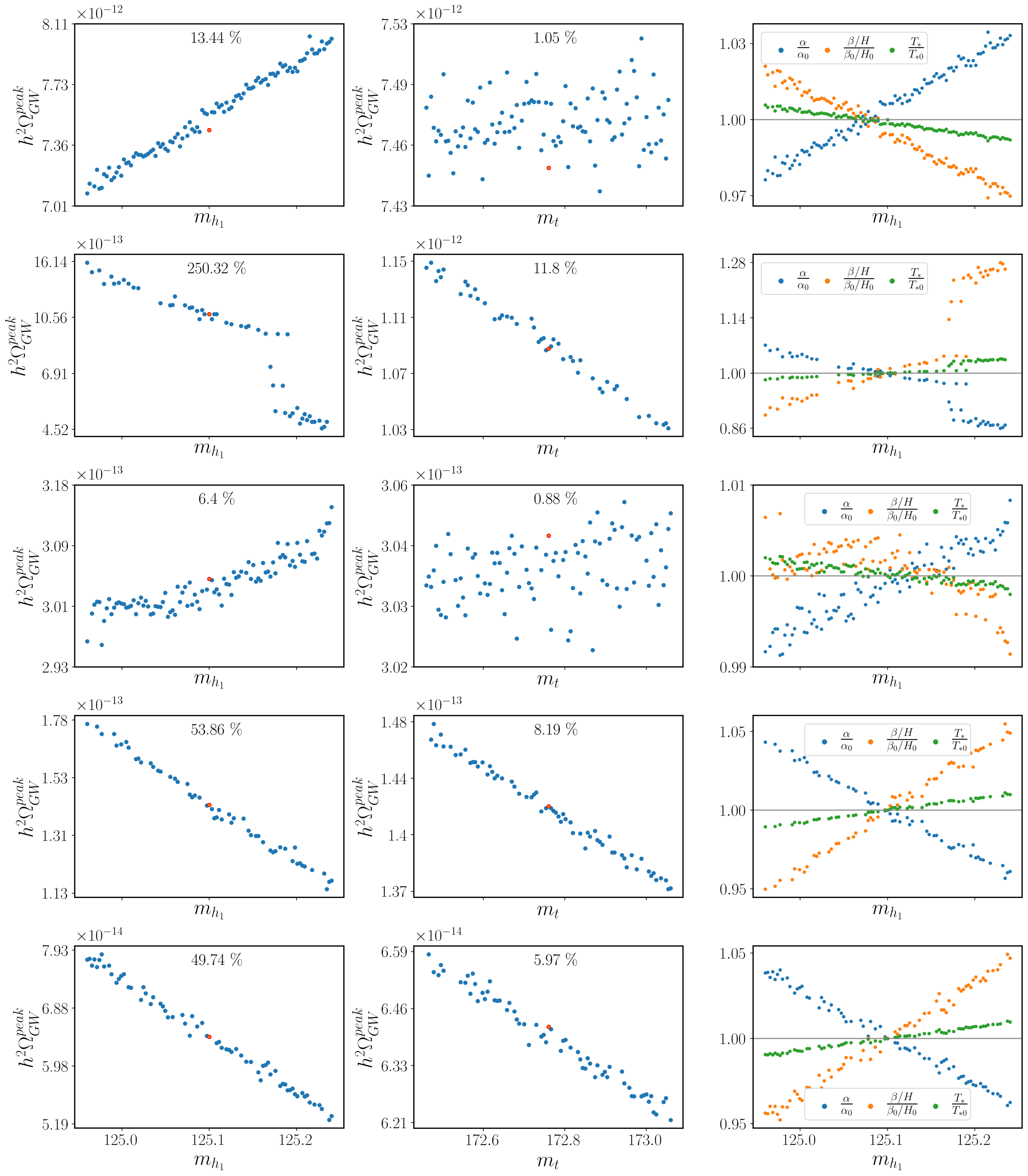}
\caption{The dependence of $h^2 \Omega^{\rm peak}_{\rm GW}$ on the Higgs boson mass $m_{h_1}$ (left) and on the top-quark mass (center). In the right plot we present the corresponding variation for the parameters $\alpha$,  $\beta/H$ and $T_*$, where the subscript $0$ denotes the central values at $m_{h_1} = 125.1$, the top text is the relative change between the strongest point and weakest point.
For each point varied all other model parameters are fixed.  The chosen original FOPT (before parameters' variation) is denoted by a red circumference. All masses are expressed in GeV.
}
\label{FigSM}
\end{figure}
  
In order to understand the impact of the variation of the SM parameters within their experimental errors we have first chosen points that are within LISA reach
taking all SM parameters with their central values according to the PDG review~\cite{Zyla:2020zbs}. We then varied each of the fermion masses, from the electron
to the top quark, the $W$ and $Z$ bosons' masses and the Higgs mass. The variation of the masses of the SM particles and the calculation of the error in the
peak of the GW power spectrum are described in detail in appendix~\ref{app:act}. We concluded that, provided we would use a smoothed action, 
the only SM masses capable of inducing a significant shift in the GW peak amplitude and frequency are the top quark mass and even more so the Higgs boson mass.

In Fig.~\ref{FigSM} we present five points that could be probed by LISA in the scope of scenario 2, identified by a red circle. In the left column we show the variation of the
peak amplitude with the variation of the Higgs mass performed from 124.96 GeV to 125.24 GeV, corresponding to an uncertainty in the Higgs mass of $\pm 1$ standard deviation, while keeping the remaining 
parameters of the SM constant together with those from the dark sector. From all the points above LISA the maximum variation found
for the peak amplitude was 250 \% for the variation with the Higgs mass.

The message is very clear and quite striking: the variation of only the Higgs mass within its experimental uncertainty leads to a variation of the peak amplitude (and also in the peak frequency) of up to 250 \% for the sample obtained. 
This could mean the following: starting with a central value giving rise to a GW signal detectable by LISA, one could move outside of the LISA sensitivity range just by varying the Higgs boson mass (within the current experimental uncertainty). Hence, the dependence on the Higgs mass is indeed meaningful and must be considered in this type of numerical studies.

The other SM parameter that can shift the GW peak amplitude and frequency is the top quark mass (middle panel in Fig.~\ref{FigSM}). The red circled points again show the central values. We have varied the top quark mass 
between 172.46 GeV and 173.06 GeV, which corresponds to an uncertainty in the top quark mass of $\pm 1$ standard deviation. 
The variation is clearly smaller than that for the Higgs mass but it is still relevant in the context of the considered variation in the top quark mass with a maximum of 50 \%. In conclusion, if a GW signal is detected one must be cautious in taking too strong conclusions about either the model or its parameters since the experimental uncertainty in the SM parameters can still play a significant role. With this in mind, a more precise determination of the Higgs and top quark masses can be rather important in light of a hypothetical discovery of a primordial GW signal and its theoretical interpretations, suggesting a further motivation for lepton colliders in the future.

A final word regarding the variation of the relevant parameters that measure the strength of the GW. Both $\alpha$ and $\beta/H$ show a shift that is proportional to the shift obtained with the SM masses with $\alpha$ with a positive slope
and $\beta/H$ with a negative one.

\subsection{Scenario 3}

\begin{figure}[h!]
\centering
\includegraphics[width = 0.80 \linewidth]{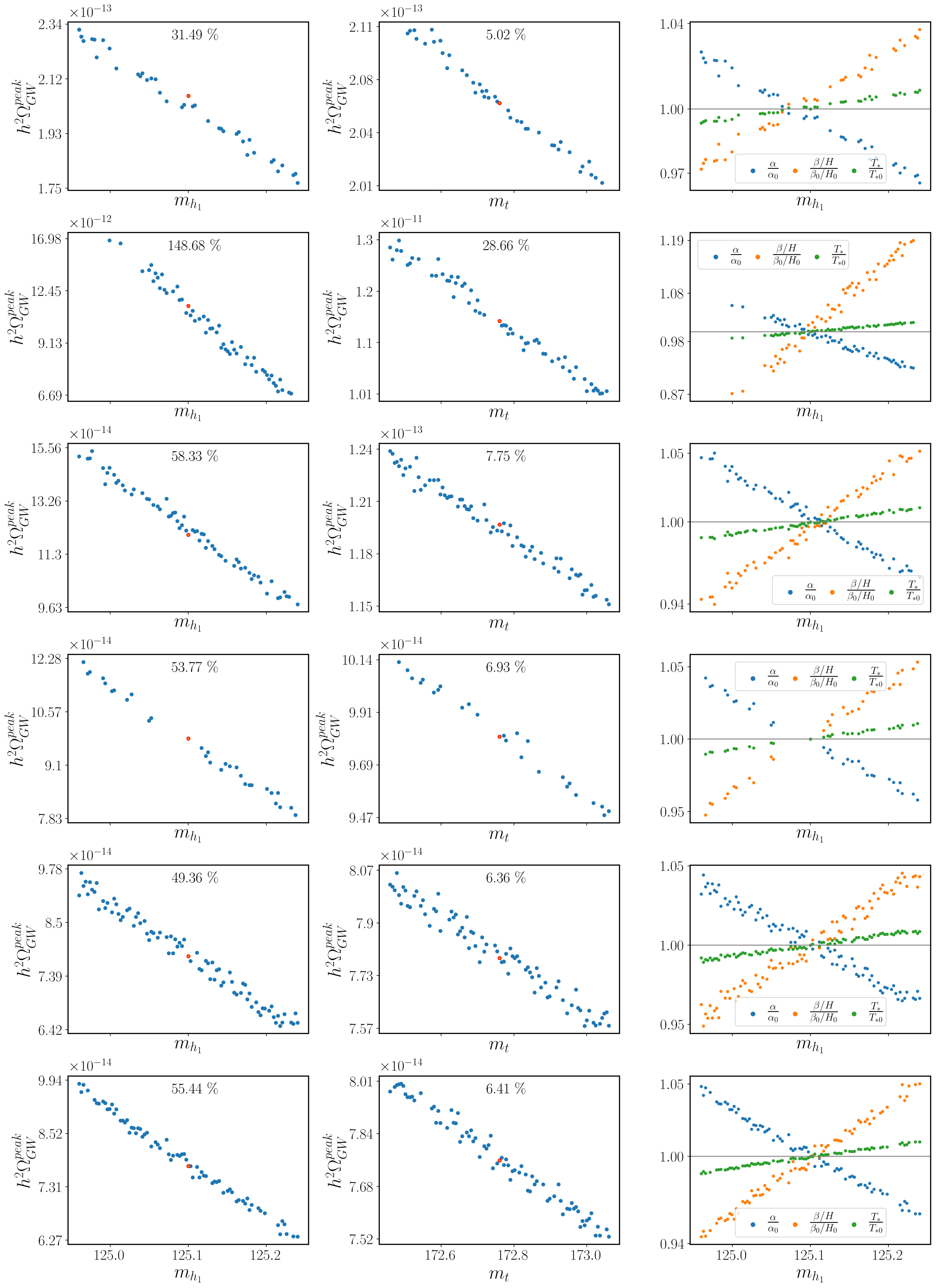}
\caption{The dependence of $h^2 \Omega^{\rm peak}_{\rm GW}$ on the Higgs boson mass $m_{h_1}$ (left) and on the top-quark mass (center). In the right plot we present the corresponding variation for the parameters $\alpha$,  $\beta/H$ and $T_*$, where the subscript $0$ denotes the central values at $m_{h_1} = 125.1$.
For each point varied all other parameters of the models are fixed. The chosen original FOPT (before parameters' variation) is denoted by a red circumference. All masses are in GeV.
}
\label{FigSc3}
\end{figure}

We now move to scenario 3. In Table~\ref{tab3} we present the ranges of variation for the input values of the model. The ranges for the parameters
of the potential were already explained. The ranges for the specific parameters of the neutrino masses and Yukawa coupling were chosen such that the Majorana-like Yukawa coupling $Y_\sigma$ is sizeable enough, but still perturbative, in order to modify the thermal coupling $c_\sigma$ in comparison to the singlet model. The parameter $M_n$ establishes a mass scale for the heavy neutrinos not to far from the EW one provided that such a scenario relies on a low-scale inverse seesaw mechanism. This is also convenient for our fixed scale treatment since renormalization group effects in the neutrino sector can be ignored.
\begin{table*}[h!]
\centering
\begin{tabular}{@{}rcccr@{}}\toprule
& \multicolumn{3}{c}{Scenario 3} \\
\cmidrule{2-4} 
& Parameter & Range & Distribution &\\ \midrule
& $m_{h_2}$ & $[50,\,1000]\,\text{GeV}$ & linear\\
& $v_{\sigma}$ & $[50,\,1000]\,\text{GeV}$ & linear &\\
& $\mu_{b}^2$ & $[-500000,\,-1250]\,\text{GeV}^2$ & linear &\\
& $\theta$ & $[-\arccos{(0.85)},\,	\arccos{(0.85)}]$ & linear &\\
& $M_n$ &  $[50,\,550]\,\text{GeV}$ & linear\\
& $Y_\sigma$ & $[0.01,\,\sqrt{4 \pi}]$ & exponential &\\
\bottomrule
\end{tabular}
\caption{Ranges of the input parameters in the scans for scenario 3.}
\label{tab3}
\end{table*}

\begin{figure}[h!]
\centering
\includegraphics[width = 0.50 \linewidth]{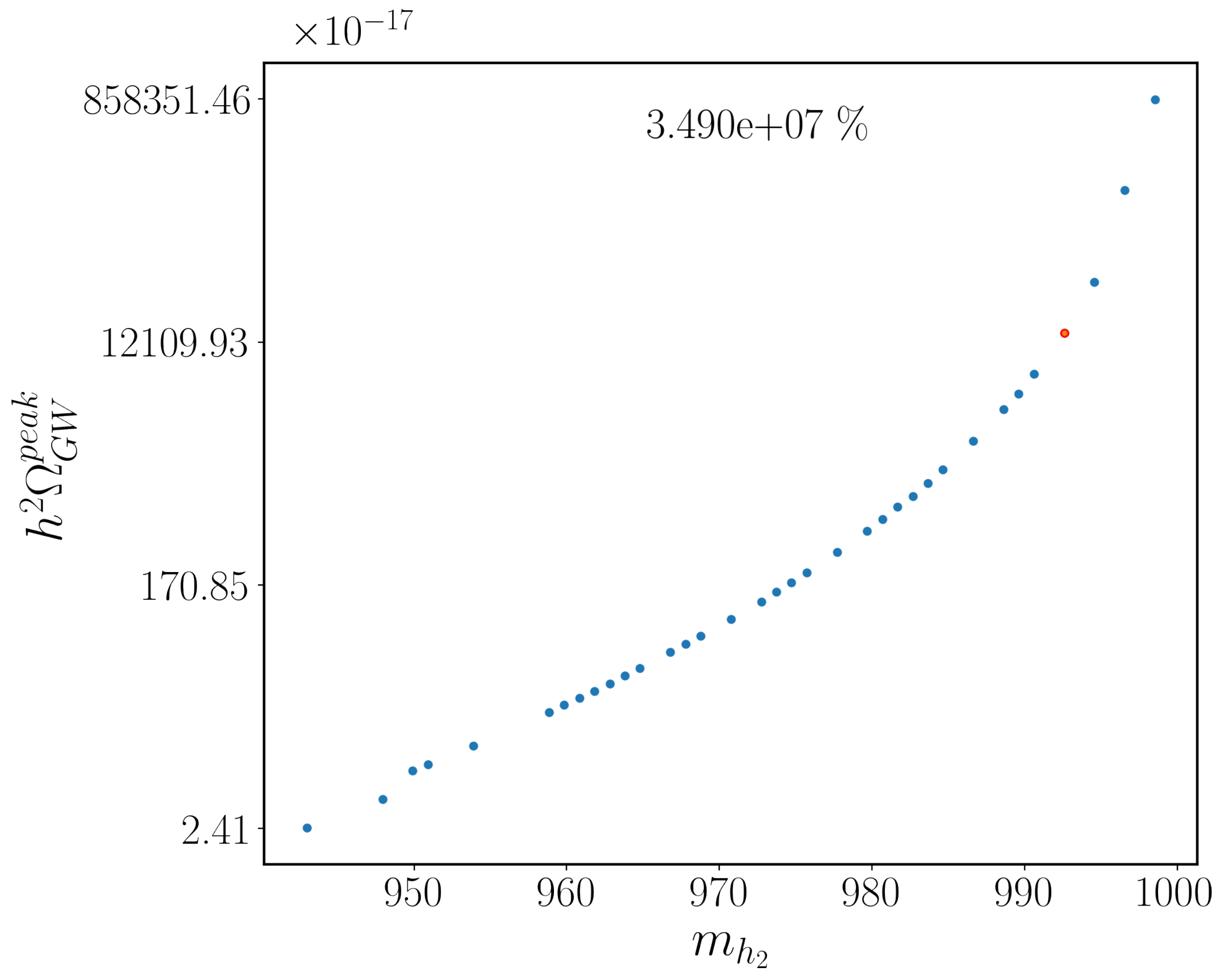}
\caption{The dependence of $h^2 \Omega^{\rm peak}_{\rm GW}$ on the second Higgs boson mass $m_{h_2}$ with all other parameters fixed.  The chosen original FOPT (before parameters' variation) is denoted by a red circumference. 
}
\label{Figmh2}
\end{figure}

In Fig.~\ref{FigSc3} we present six points that could be probed by LISA in the scope of scenario 3, identified by a red circle. As for scenario 2, we present in the left column the variation of the
peak amplitude with the variation of the Higgs mass in the interval 124.96 GeV to 125.24 GeV ( $\pm 1$ standard deviation), while keeping the remaining 
parameters of the SM constant together with those from the dark sector. Again the top quark mass (middle column) was varied between 172.46 GeV and 173.06 GeV. 
From all the points above LISA the maximum variation found
for the peak amplitude was 150 \% for the variation with the Higgs mass and 30 \% for the variation with the top quark mass. This is in line with what was obtained for scenario 2.

We end this section showing  in Fig.~\ref{Figmh2} the dependence of $h^2 \Omega^{\rm peak}_{\rm GW}$ on the second Higgs boson mass $m_{h_2}$ with all other parameters fixed. As expected, since we allow for a much larger variation
of order 5\%, the impact on the amplitude of the primordial GW spectrum is huge spanning several orders of magnitude.
This is the general trend for all other points in the scan with strong FOPTs. 

\section{Comparison of the results for three scenarios}
\label{sec:3Sc}

In this last section we will focus in more detail on the observability of GWs by the LISA experiment starting with scenario 1. In Table~\ref{tab1} we show the range of variation for the parameters in scenario1.
\begin{table*}[h!]
\centering
\begin{tabular}{@{}rcccr@{}}\toprule
& \multicolumn{3}{c}{Scenario 1} \\
\cmidrule{2-4} 
& Parameter & Range & Distribution &\\ \midrule
& $m_\text{D1}$ & $[50,\,1000]\,\text{GeV}$ & linear\\
& $m_\text{D2}$ & $[50,\,1000]\,\text{GeV}$ & linear &\\
& $\lambda_{\Phi \sigma}$ & $\pm[0.05,\,1]$ & exponential &\\
& $\lambda_{\sigma}$ & $[0.05,\,1]$ & exponential &\\
\bottomrule
\end{tabular}
\caption{Ranges of the input parameters in the scans for scenario 1.}
\label{tab1}
\end{table*}
\begin{figure}[h!]
\centering
\includegraphics[width = 0.9 \linewidth]{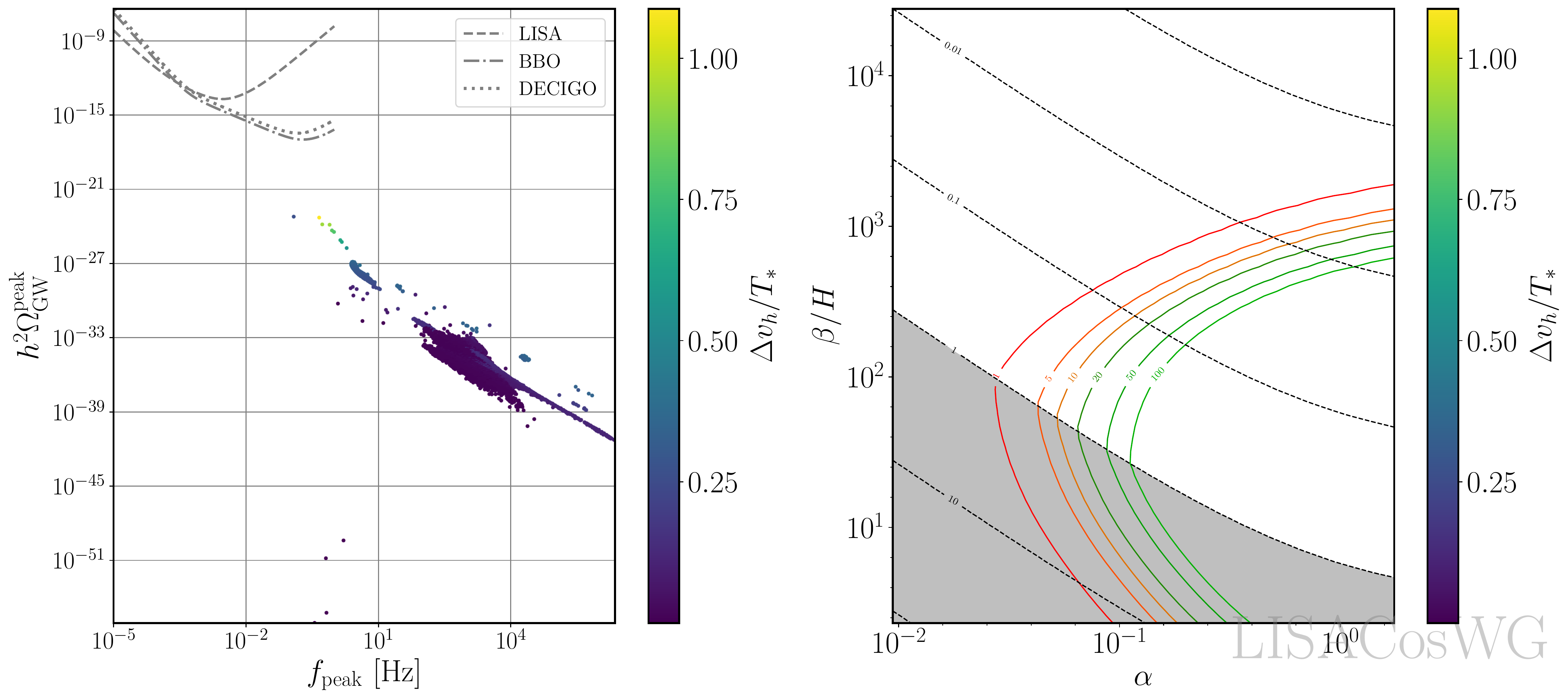}
\caption{Scatter plots showing the typical strength of the phase transitions $\Delta v_{h}/T_*$ in the colour scale of each plot. On the left panel the position of the GW peak is shown whereas on the right we show the corresponding SNR for a mission profile of three years. The colored lines show the SNR that depends on $T_{*}$, $g_{\ast}(T_{*})$ and $v_{\text{b}}$. The
dotted curves are contour lines representing the shock formation time $\tau_{\text{sh}}$ as defined in \eqref{eq:Opeak2}. The grey shaded region corresponds to an acoustic period lasting longer than a Hubble time and it is where the sound waves treatment is mostly reliable ~\protect\cite{Hindmarsh:2017gnf, Ellis:2018mja}.}
\label{novev}
\end{figure}

\begin{figure}[h!]
\centering
\includegraphics[width = 0.9 \linewidth]{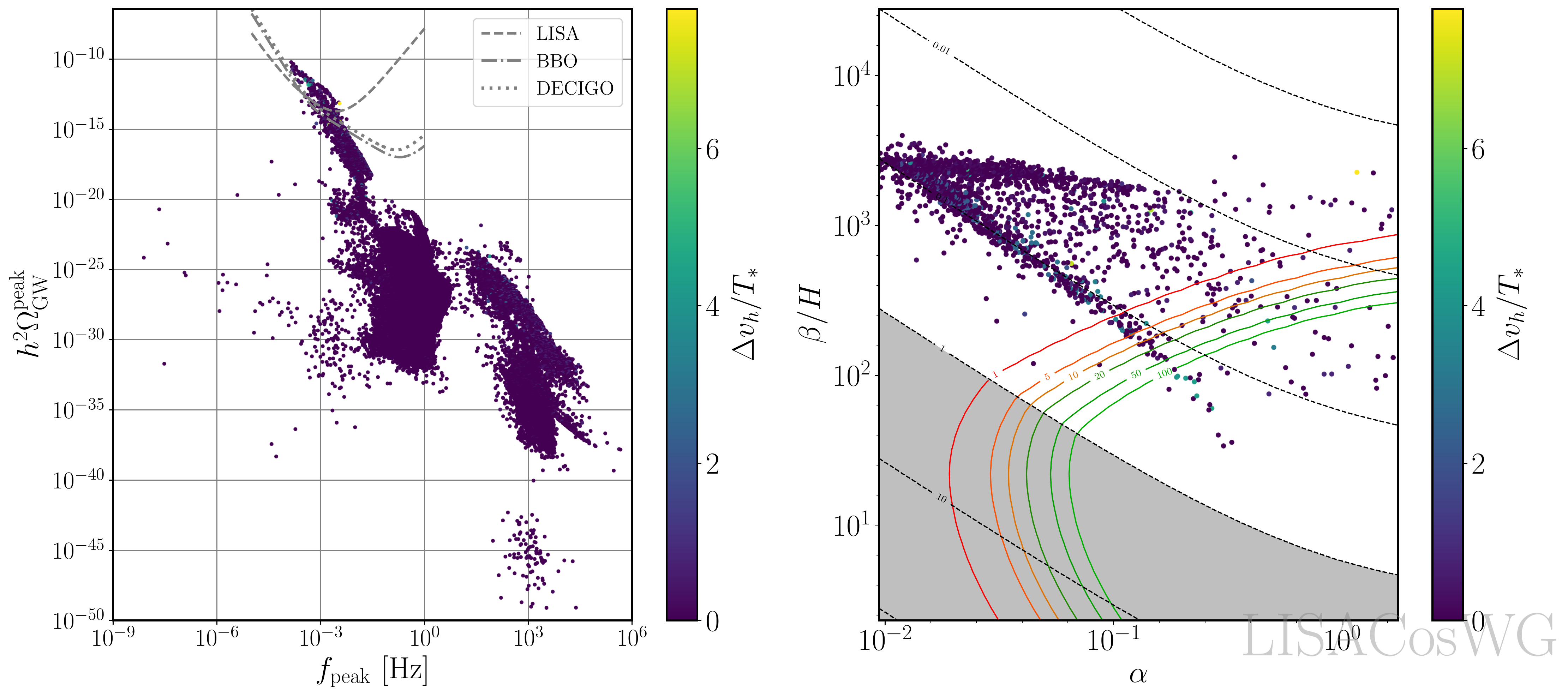}
\caption{Scatter plots showing the typical strength of the phase transitions $\Delta v_{h}/T_*$ and SNR for scenario 2.}
\label{withvev}
\end{figure}

In Fig.~\ref{novev} we present scatter plots for scenario 1  showing the GW peak position as a function of the strength of the phase transitions $\Delta v_{h}/T_*$ in the colour scale (left) and the corresponding signal-to-noise (SNR) ratio for the phase transition (right) for a mission profile of 3 years. The color grade scale is the same on both plots. The right panel was generated using \texttt{PTPlot 1.0.1}~\cite{Caprini:2019egz}. The colored isolines display the expected values for the SNR that depend on $T_*$, $g_\ast$ and $v_b$ while the dashed black contour lines represent the shock formation time $\tau_\mathrm{sh}$ (see \eqref{eq:Opeak2}). The grey shaded region corresponds to an acoustic period lasting longer than a Hubble time and it is where the sound waves treatment is mostly reliable~\cite{Ellis:2018mja,Hindmarsh:2017gnf}. For $\tau_\mathrm{sh} \ll 1$, the turbulence effects may become important dumping the acoustic contribution. However, none of our points feature a too small shock formation time. Using the formula for turbulence effects in~Refs.~\cite{Caprini:2009yp,Ellis:2018mja} for an estimate, we realize that it does indeed have very little impact in the peak position on the left panel.  This first plot is just shown for reference. As concluded in the previous sections in the case where the singlet does not acquire a VEV, phase transitions are very weak with peaks amplitudes below $10^{-22}$. 
We have generated a total of 73 047 points points with a FOPT and several dedicated scans were performed but the trend did not change. 
Note that the actual role of turbulence is not yet well understood~\cite{Kosowsky:2001xp,Gogoberidze:2007an,Niksa:2018ofa} and further studies are needed for a more reliable calculation of such a component. The SNR contours on the right panel take into account the effect of an increasingly short-lasting shock formation at the cost of a decreasing SNR value.

In Fig.~\ref{withvev} we present a similar scatter plot but now for scenario 2.  
As previously discussed there are good chances of probing this model in some regions of the parameter space. In particular, we have found 94 points with SNR larger than 5 out of which 45 feature a SNR above 100.
  
\begin{figure}[h!]
\centering
\includegraphics[width = 0.9 \linewidth]{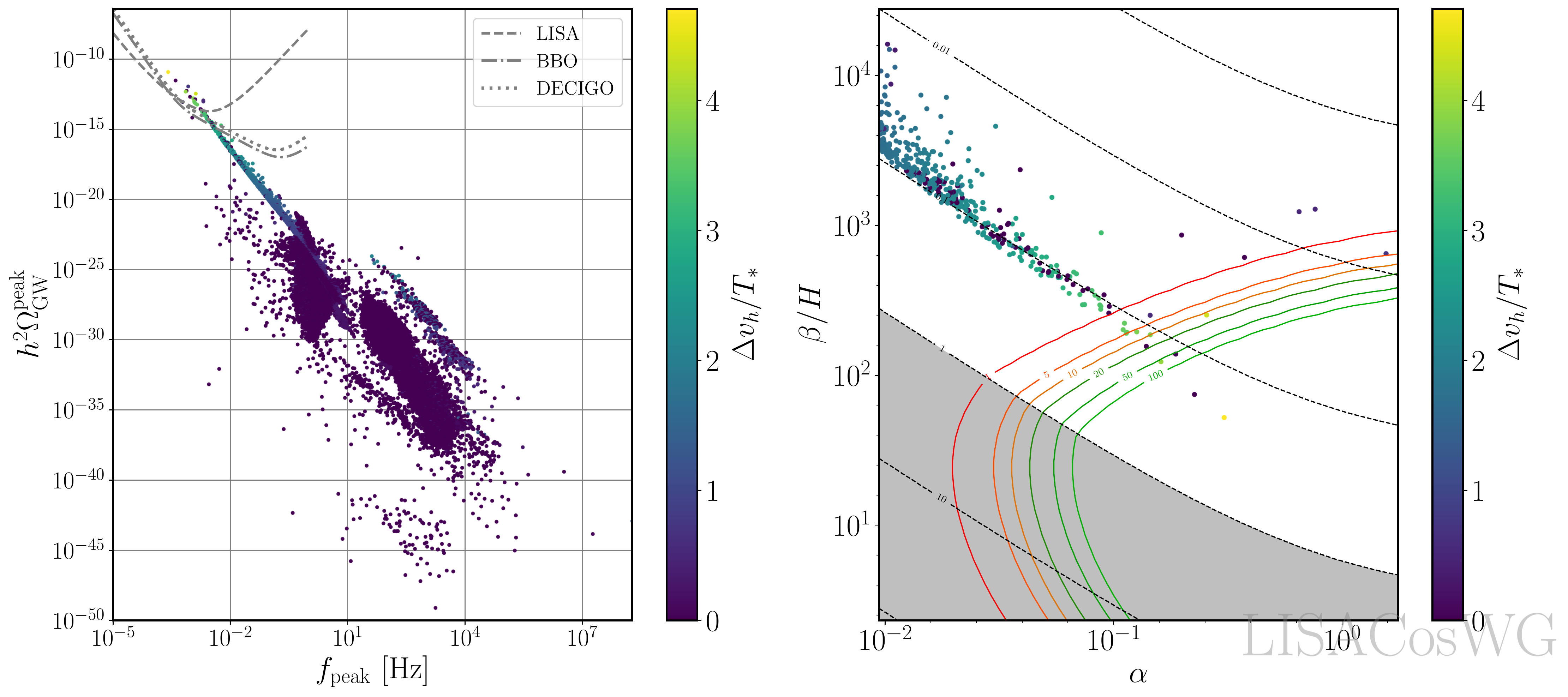}
\caption{Scatter plots showing the typical strength of the phase transitions $\Delta v_{h}/T_*$ and SNR for scenario 3.}
\label{withneutrino}
\end{figure}

Finally in Fig.~\ref{withneutrino} we show similar plots but for scenario 3. We found that there was no major difference with the addition of right-handed Majorana neutrinos to the broken phase. 
We again find 8 points with SNR larger than 5 and two points withs SNR above 100. The sample is 42 509. 
Further studies of Majoron models in the context of primordial GWs were performed in~Ref.~\cite{Addazi:2019dqt} and, more recently, in~Ref.~\cite{DiBari:2021dri}.

\section{Conclusions}
\label{sec:Conc}

We have discussed the observation of primordial GWs originating from a strong FOPT using as a benchmark model an extension of the SM by a complex singlet scalar field.
Three different scenarios were proposed: scenario 1 (no zero temperature VEV in the singlet) leading to the SM plus two DM candidates that only communicate via the portal
coupling; scenario 2 (zero temperature VEV in the real part of the singlet) leading to a mixing between the two CP-even scalars plus one DM candidate; finally scenario 3
where besides the singlet field, which also acquires a zero temperature finite VEV, six heavy non-SM  neutrinos are added to the model.

Scenario 2 was then used to answer some key questions that relate the Higgs potential with the detection of GWs.
Our conclusion is that the variation of the SM particle masses does indeed lead to sizeable differences both in the peak of the GW power spectrum
and on its frequency. This is particularly true for the Higgs mass but also for the top quark mass. The other SM masses have a very mild, if any,
impact on the detection of GW in the near future. We should underline the point that the variation of the Higgs mass within the measured experimental
error can lead to at least 250 \% magnitude change in the GW peak. It is then crucial to increase the precision in the Higgs mass measurement if this effect is to be mitigated, 
which can be seen as a motivation for lepton collider machines. Note that this is the conclusion for a particular very simple model and that there could be models with even more dramatic changes. 
After concluding that the SM parameters have an impact on the detection of GWs, it does not come as a surprise that the other parameters from the potential, including the DM masses and the portal coupling also play an important role. 
The addition of new particles to the version of the model with spontaneously broken global $U(1)$ does not make any dramatic change in the results regarding the variation with the SM parameters. 

The second question we wanted to answer was if the spontaneous symmetry breaking of a given model would lead to relevant differences in the strength of the GW spectrum. We concluded
that indeed it does. Taking the exact same model, with a phase with two DM candidates and going to a phase with one DM candidate (as the singlet VEV breaks the symmetry) the
chances of detecting GWs by LISA (or even other future experiments) go from negligible to excellent. 
Hence, we believe that our study delivers two clear messages. First, any such study needs to take into account the precision in the measurements of at least the Higgs mass and the top quark mass. Second,
even the same model, if considered in different phases at zero temperature, can exhibit a very distinct behaviour in what concerns the detection of GWs originating from strong FOPTs. 
Finally, we have discussed technical issues in $\beta/H$ computations and how not taking proper care of the derivative of the action can lead to numerical instabilities and, hence, to wrong results.

\appendix

\section{Smoothing the action}
\label{app:act}

The value of ${\beta}/{H}$ is given by Eq. (\ref{betaH}). As can be seen from this expression, the calculation involves the derivative of the action with respect to the temperature. A possible method to calculate it numerically is to use the Difference Quotient Method (DQM) 
with the bounce action numerically computed by \texttt{CosmoTransitions} as
\begin{equation}
\frac{\beta}{H} = T_*  \left. \frac{d}{d T} \left( \frac{\hat{S}_3(T)}{T}\right) \right|_{T_*} \approx T_* \frac{1}{2 \cdot\Delta T}\left(\left( \frac{\hat{S}_3(T)}{T}\right)\Bigg|_{T = T_*+\Delta T}-\left( \frac{\hat{S}_3(T)}{T}\right)\Bigg|_{T = T_*-\Delta T}\right)\,,
\end{equation}
where $\Delta T$ is the small step of the DQM. This method correctly calculates ${\beta}/{H}$ for points with a strong GW signal, but for weaker points, numerical errors in the calculation of $\hat{S}_3$ make the DQM not completely reliable.
Our solution is to interpolate the action around $T_n$ starting by sampling $N = \{60\,,75\,,90\,,105\}$ bounce actions\footnote{\texttt{CosmoTransitions} is not always able to calculate all of the $N$ bounce actions. When this happens we do the fit with whichever points it managed to calculate.} inside the interval ranging from $\max\{T_n-30,T_n/2\} \text{ GeV}$, which should leave enough room to calculate the percolation temperature, up to $T_c-3\text{ GeV}$. We do not interpolate exactly up to $T_c$ to prevent numerical instabilities regarding the existence/location of the minimum. Moreover, for the calculation of the percolation temperature, $T_*$, this truncation has a negligible effect since the biggest contribution comes from the epoch around $T_n$. 
The $4$ independent samples of $N$ actions are linearly distributed inside the mentioned interval, in each sample. If we calculate more than $7$ bounce actions then a degree 6 polynomial in $T$ can be fitted that models $\hat{S}_3/T$. After this procedure the calculation of the derivative is trivial. 
Applying this method to four independent samples of points allows us to calculate $\beta/H$ four times. We consider the most correct value for $\beta/H$ to be the average between all 4 samples. This allows us to estimate an error $\Delta\left(\beta/H\right)$ for our method which we define as the difference between the biggest $\beta/H$ and the smallest $\beta/H$ divided by two,
\begin{align}
\Delta\left(\beta/H\right) = \frac{\text{max}\{\beta/H\}-\text{min}\{\beta/H\}}{2}\,.
\end{align}

\begin{figure}[h!]
\centering
\includegraphics[width = 0.32 \linewidth]{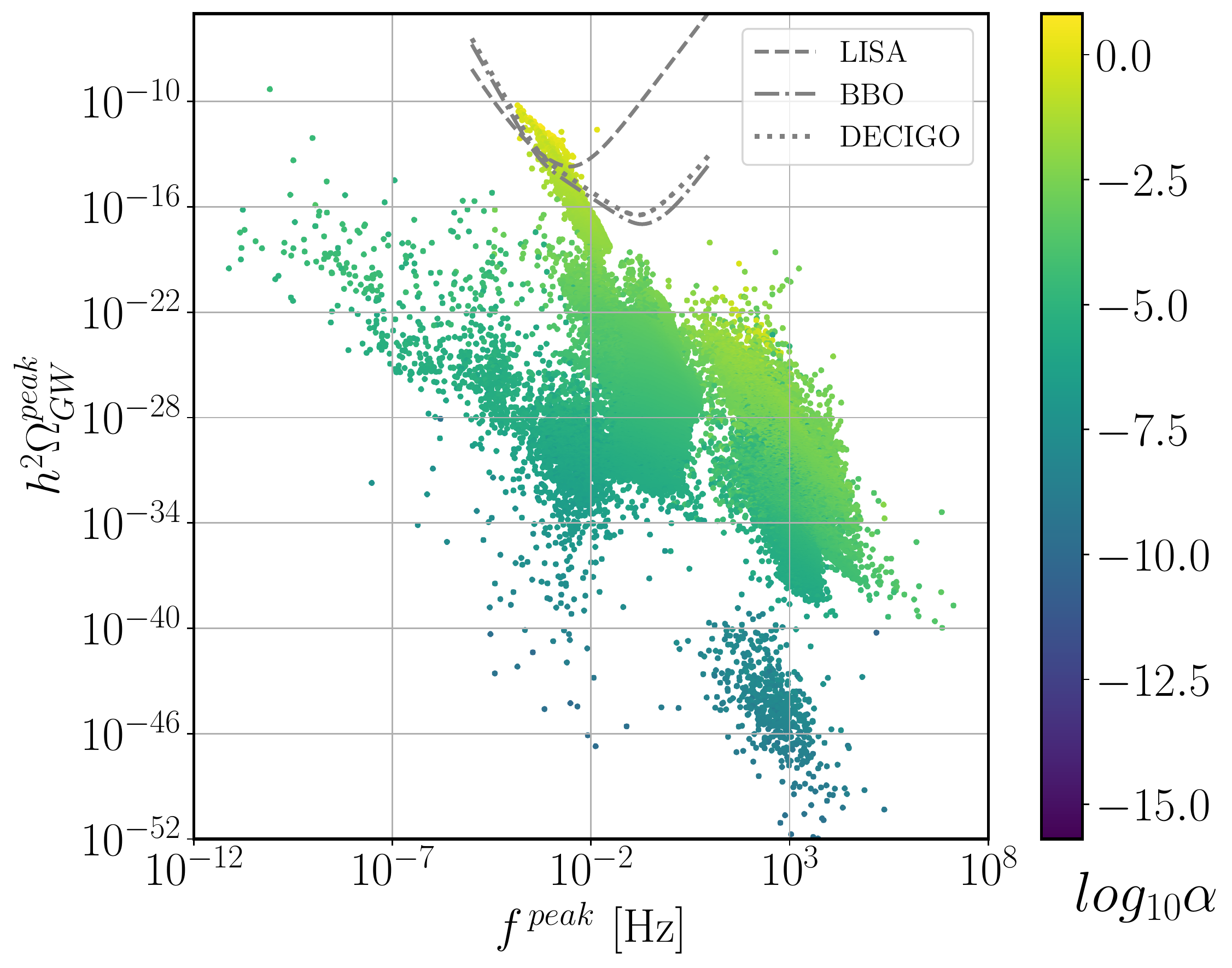}
\includegraphics[width = 0.32 \linewidth]{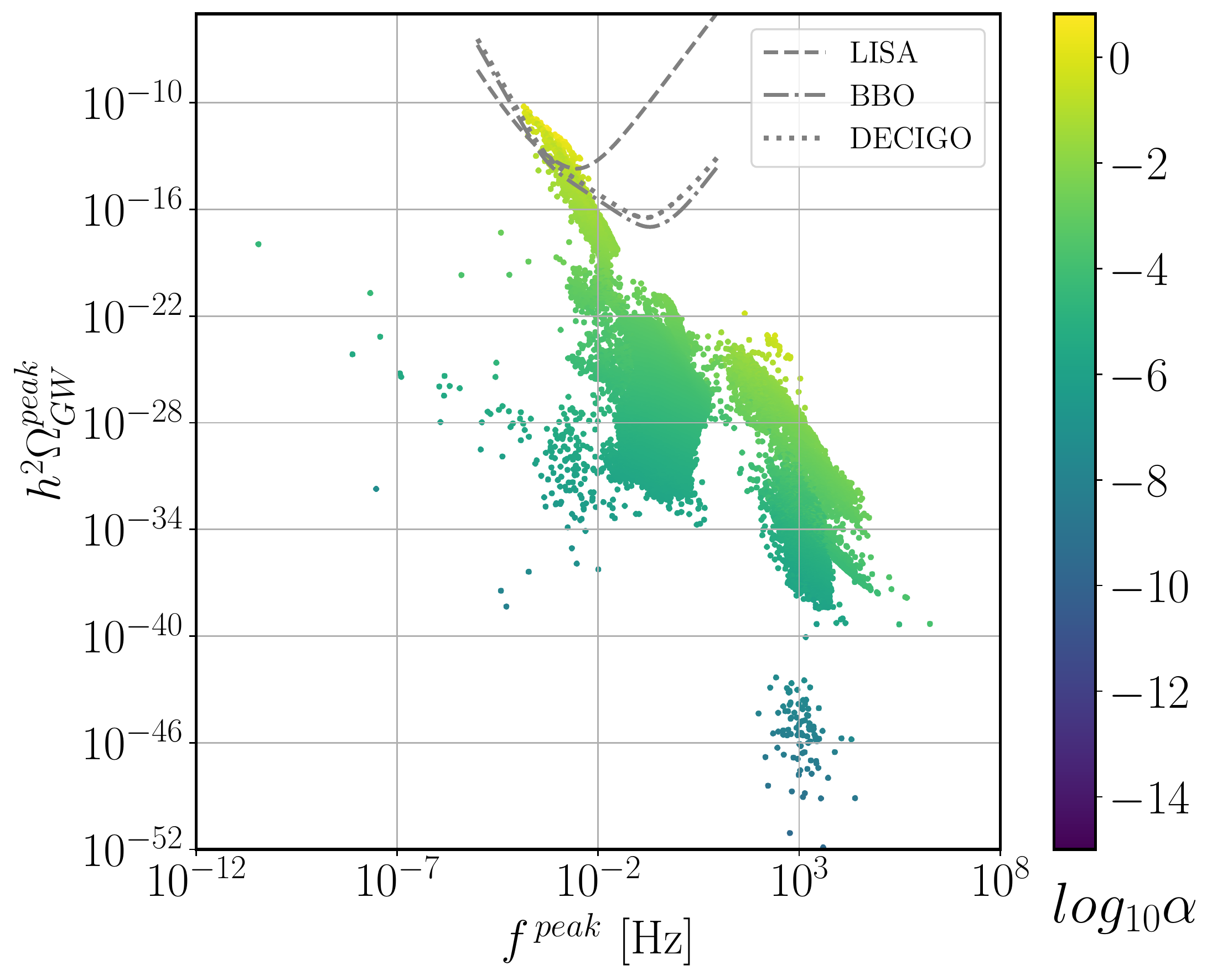}
\includegraphics[width = 0.32 \linewidth]{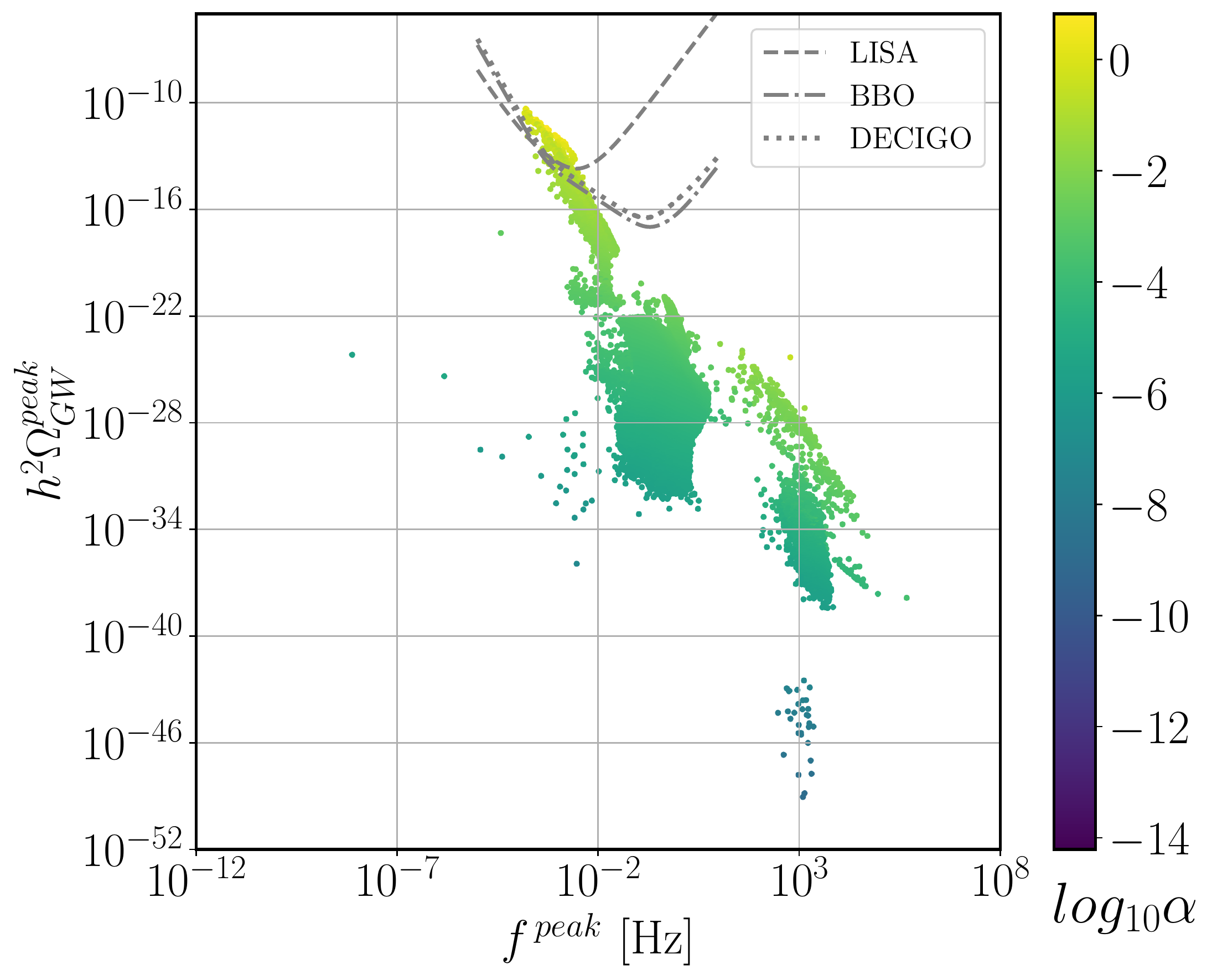}
\caption{The peak-amplitude for the GW signal $h^2 \Omega^{\rm peak}_{\rm GW}$ as a function of the peak frequency $f_{\rm peak}$ in logarithmic scale for scenario 2.
The scatter plots present, in the colour bar, the strength of the phase transition $\alpha$. In the left plot, there are no restrictions related to the calculation of $\beta/H$, 
in the middle plot only points with $\Delta (\beta/H) < 0.25$ are accepted, and in the right panel only points with  $\Delta (\beta/H) < 0.05$ are accepted.}
\label{Alpha_percent}
\end{figure}

The next question to ask is: which points will we consider as valid when we use this method on the output of \texttt{CosmoTransitions}?
In Fig.~\ref{Alpha_percent} we now present a scatter plot with the GW signal $h^2 \Omega^{\rm peak}_{\rm GW}$ as a function of the peak frequency $f^{\rm peak}$ for three different levels of constraints set upon the uncertainty of $\beta/H$. 
On the left panel, a total of 215 065 points are shown
which are the ones with no restrictions on $\Delta (\beta/H)$, that is, the original set of points before the smoothing procedure is applied. Once we impose that $\Delta (\beta/H) < 0.25$ (middle panel) the number of points is reduced to 192 316 and if we further restrict the error  $\Delta (\beta/H) < 0.05$,
the number of allowed points is reduced to 108 032, that is, only 50\% of the points remain. In the plots presented in the paper, all points have $\Delta (\beta/H) < 0.25$. We did not want to further restrict the error because it could be that we were also losing too many good points.

Our method has another benefit, the interpolation of the bounce action provides us with an approximate analytical expression for $\hat{S}_3/T$ which, in turn, gives us an approximate analytical expression for the tunnelling rate $\Gamma(T)$. This allows us to promptly calculate the nucleation and percolation temperatures and because we have 4 samples we can also estimate the error associated with our calculation of the characteristic temperatures.

\newpage
\subsubsection*{Acknowledgments}
JV and RS are supported by Centro de Física Teórica e Computacional da Universidade de Lisboa (CFTC-UL) through the Portuguese Foundation for Science and Technology (FCT), 
under Contracts UIDB/00618/2020, UIDP/00618/2020. JV and RS are also supported by the FCT projects PTDC/FIS-PAR/31000/2017,  CERN/FIS-PAR/0014/2019, and by the National Science Centre, 
Poland, the HARMONIA project under contract UMO- 2015/18/M/ST2/00518 (2016-2021). FFF, and APM are supported by the Center for Research and Development in Mathematics and Applications (CIDMA) through 
FCT with
references UIDB/04106/2020 and UIDP/04106/2020. They are also supported by the projects 
PTDC/FIS-PAR/31000/2017, CERN/FIS-PAR/0002/2019 and PTDC/FIS-AST/3041/2020. APM is also supported by national funds (OE), through FCT, I.P., in the scope of the framework 
contract foreseen in the numbers 4, 5 and 6 of the article 23, of the Decree-Law 57/2016, of August 29, changed by Law 57/2017, of July 19. GL, AN and JO thank CTFC-UL for 
the opportunity to participate in the Summer Internships 2020 program where this work has started. RP~is supported in part by the Swedish Research Council grant, contract 
number 2016-05996, as well as by the European Research Council (ERC) under the European Union's Horizon 2020 research and innovation programme (grant agreement No 668679).
The authors acknowledge the support of the FCT Advanced Computing Project that provided computational resources via the project CPCA/A00/7395/2020 and that 
of INCD funded by FCT and FEDER, project 01/SAICT/2016 nº 022153. The authors also acknowledge the use of the computer resources at Artemisa, 
funded by the European Union ERDF and Comunitat Valenciana as well as the technical support provided by the Instituto de Física Corpuscular, IFIC (CSIC-UV).

\bibliographystyle{bib-style}
\bibliography{GW.bib}

\end{document}